\documentclass[twocolumn]{aastex631}

\usepackage{savesym}
\savesymbol{tablenum}
\usepackage[sicmds,freestanding]{hepunits}
\restoresymbol{SIX}{tablenum}
\usepackage{amssymb,mathtools,bm}
\usepackage{graphicx, color}
\usepackage[dvipsnames]{xcolor}
\usepackage{needspace}
\usepackage{filecontents}
\usepackage{multirow}
 \usepackage{hyperref} 
\hypersetup{
    colorlinks=true,       
    linkcolor=blue,        
    citecolor=blue,        
    filecolor=magenta,     
    urlcolor=blue          
}

\usepackage{slashed}

\usepackage{url}
\usepackage[utf8]{inputenc}
\usepackage[english]{babel}
\usepackage{mathrsfs}
\usepackage{rotating}

\newlength{\dhatheight}

\newcommand{\osim}{\ensuremath{\mathord{\sim}}}
\newcommand{\Msun}{\ensuremath{\mathrm{M}_{\odot}}}
\newcommand{\Lsun}{\ensuremath{\mathrm{L}_{\odot}}}
\newcommand{\sigmaLOS}{\ensuremath{\sigma_{\mathrm{\scriptscriptstyle LOS}}}}
\newcommand{\Mstar}{\ensuremath{M_{\star}}}
\newcommand{\Mpeak}{\ensuremath{M_\mathrm{peak}}}
\newcommand{\Mvir}{\ensuremath{M_\mathrm{vir}}}
\newcommand{\Vcirc}{\ensuremath{V_\mathrm{circ}}}
\newcommand{\Mhalf}{\ensuremath{M_{1/2}}}
\newcommand{\rhalf}{\ensuremath{r_{1/2}}}
\newcommand{\vmax}{\ensuremath{v_\mathrm{max}}}
\newcommand{\rmax}{\ensuremath{r_\mathrm{max}}}
\newcommand{\concmass}{\ensuremath{c-M}}
\newcommand{\kms}{\ensuremath{\mathrm{km}~\mathrm{s}^{-1}}}

\newcommand{\SatGen}{\texttt{SatGen}}

\defcitealias{Part_II}{Paper~II}
\newcommand{\papertwo}{\citetalias{Part_II}}
\defcitealias{Fattahi:2018ioj}{Fattahi+18}
\newcommand{\Fattahi}{\citetalias{Fattahi:2018ioj}}
\defcitealias{2024arXiv240815214K}{Kim+24}
\newcommand{\Kim}{\citetalias{2024arXiv240815214K}}
\defcitealias{2023ApJ...956....6D}{Danieli+23}
\newcommand{\Danieli}{\citetalias{2023ApJ...956....6D}}
\defcitealias{2018MNRAS.477.1822M}{Moster+18}
\newcommand{\Moster}{\citetalias{2018MNRAS.477.1822M}}
\defcitealias{Wolf:2009tu}{Wolf+10}
\newcommand{\Wolf}{\citetalias{Wolf:2009tu}}
\defcitealias{Errani_2018}{Errani+18}
\newcommand{\Errani}{\citetalias{Errani_2018}}
\defcitealias{Diemer:2018vmz}{Diemer+19}
\newcommand{\Diemer}{\citetalias{Diemer:2018vmz}}
\defcitealias{Zhao:2008wd}{Zhao+09}
\newcommand{\Zhao}{\citetalias{Zhao:2008wd}}

\newcommand{\es}[2] {\begin{equation} \label{#1} \begin{split} #2 \end{split} \end{equation}}

\makeatletter
\let\frontmatter@title@above=\relax
\makeatother

\begin{document}

\input{aastex-deluxetable-spacing-fix}

\title{
Semi-analytic Inference of Satellite Densities in the Cold Dark Matter Model\\ \vspace{0.05in}
{\small Part I. Comparison to Ultra-faint Dwarf Kinematics}
}

\author{Kailash Raman}
\affiliation{Leinweber Institute for Theoretical Physics, University of California, Berkeley, CA 94720, U.S.A.}
\affiliation{Theoretical Physics Group, Lawrence Berkeley National Laboratory, Berkeley, CA 94720, U.S.A.}

\author{Dylan Folsom}
\affiliation{Department of Physics, Princeton University, Princeton, NJ 08544, U.S.A.}

\author{Manoj Kaplinghat}
\affiliation{Department of Physics and Astronomy, University of California - Irvine, Irvine, CA 92697, U.S.A.}

\author{Mariangela Lisanti}
\affiliation{Department of Physics, Princeton University, Princeton, NJ 08544, U.S.A.}
\affiliation{Center for Computational Astrophysics, Flatiron Institute, New York, NY 10010, U.S.A.}

\author{Benjamin R. Safdi}
\affiliation{Leinweber Institute for Theoretical Physics, University of California, Berkeley, CA 94720, U.S.A.}
\affiliation{Theoretical Physics Group, Lawrence Berkeley National Laboratory, Berkeley, CA 94720, U.S.A.}

\date{\today}

\begin{abstract}
Ultra-faint dwarf galaxies are critical testing grounds for probing the limits of galaxy formation in the cold dark matter~(CDM) paradigm.
Employing a semi-analytic cosmological satellite generator that captures the expected CDM halo population, we estimate Milky Way dwarf density profiles through two methods: a kinematic approach using stellar velocity dispersions and a separate method based on dwarf stellar masses. The kinematic approach yields a larger diversity in central dark matter densities than expected from the CDM population, as inferred from the stellar-to-halo mass relation, with compact ultra-faints appearing overdense and larger systems appearing underdense. For the ultra-faint dwarfs with at least 10 stars with spectroscopic measurements, this discrepancy persists at the $\osim2.4\sigma$ level across all considered systematic variations on the semi-analytic modeling. Our framework introduces a novel, robust procedure for testing the consistency of the observed population of Milky Way satellites with cosmological expectations.  As future surveys discover new dwarf satellites and refine stellar velocity measurements, updating this analysis will provide an increasingly stringent test of the CDM paradigm. 
\end{abstract} 

\Needspace{4\baselineskip}
\section{Introduction}

Milky Way~(MW) satellite galaxies are among the most powerful probes currently available for studying the particle nature of dark matter~(DM). The ultra-faint dwarfs~(UFDs), defined here as galaxies with a stellar mass of $\Mstar{}<10^5~\Msun{}$, are excellent environments for testing galaxy formation in a cold DM~(CDM) context. 
We introduce a semi-analytic framework to model the density distributions of these dwarf systems and present the results in a two-part series. This paper, the first in the series, introduces the modeling approach and presents a framework for testing the consistency of the observed MW satellite population against CDM predictions.  The second paper~\citep[][hereafter \papertwo{}]{Part_II} discusses the implications of these inferred densities to searches for DM annihilation and decay.

The population of known UFDs has grown dramatically over the past two decades. The first UFDs were discovered as stellar overdensities in the Sloan Digital Sky Survey \citep{Willman:2004kk,SDSS:2006fpg}, and spectroscopic follow-up with Keck/DEIMOS revealed that they are the most DM-dominated galaxies known, with mass-to-light ratios approaching $\osim 10^3\;\Msun{}/\Lsun{}$~\citep{Simon:2007dq}. The Dark Energy Survey subsequently uncovered a further wave of southern-hemisphere satellites~\citep{2015ApJ...807...50B,Fermi-LAT:2015ycq}, and targeted searches continue to expand the census~\citep[see, e.g., the review by][]{2019ARA&A..57..375S}. Characterizing these systems requires spectroscopic measurements of individual stellar velocities, which remain challenging given their sparse stellar populations. Two recent efforts have substantially updated the observational landscape: the Local Volume Database (LVDB;~\citealt{Pace:2024sys}), which provides a uniform compilation of dwarf galaxy properties, and the Keck/DEIMOS Stellar Archive (KDSA;~\citealt{geha2026keckdeimosstellararchivei,geha2026keckdeimosstellararchiveii}), which presents a reanalysis of spectroscopic data for over 70 MW satellites with homogeneous velocity and metallicity measurements. The data from these compilations underpin the analyses in this work.

The DM density profiles of UFDs are commonly inferred using Jeans modeling~\citep{1915MNRAS..76...70J,1980MNRAS.190..873B,1985AJ.....90.1027M,1992ApJ...391..531D}---see, e.g.,~\citet{Strigari:2006rd,Martinez:2010xn,Pace:2018tin,Bonnivard:2015xpq,Bonnivard:2014kza,Geringer-Sameth:2014yza,Chang:2020rem} for some recent applications to MW UFDs.  This procedure describes the phase-space density of tracer stars (typically assumed to be in equilibrium) in the gravitational potential of the dwarf (typically assumed to be spherically symmetric) with the time-independent collisionless Boltzmann equation.  Integrating this equation over velocities yields the Jeans equation, which relates the gravitational potential to the velocity dispersion and stellar density of the system. By measuring projections of the velocity dispersion, for example along the line of sight~(LOS), and the stellar density distribution, one may infer model parameters describing the gravitational potential through, e.g., Bayesian techniques. While Jeans modeling can accurately recover DM density profiles for larger dwarf galaxies with extensive stellar kinematic data, the results for the inner regions of UFDs are sensitive to statistical priors given the low number of available stellar kinematic samples~\citep{Chang:2020rem}. Additionally, for compact UFDs, the lack of stars in the outer regions of the galaxy may hamper the ability to capture the tidal stripping of the DM halo.

Other techniques have also been used to infer DM halo properties from observed stellar kinematics in the MW dwarfs. Schwarzschild orbit-superposition modeling has been applied to several classical dwarfs, including Fornax, Draco, Sculptor, Carina, and Sextans~\citep{2012ApJ...746...89J,2013ApJ...763...91J,2013ApJ...775L..30J,2013MNRAS.433.3173B,Breddels:2013dga}. Phase-space distribution function modeling, which parameterizes the stellar distribution function directly, has likewise been used to constrain DM profiles in MW dwarfs~\citep{2011MNRAS.411.2118A,2012MNRAS.419..184A,Strigari:2014yea,2019MNRAS.488.2423P,2024MNRAS.532.4157A}. More recently, \citet{Nguyen:2026gap} applied neural posterior estimation to estimate the DM profiles of two dwarf galaxies, Draco and Bo\"otes I. Similar to Jeans analyses, these approaches aim to infer the DM halo profiles consistent with the observed kinematic data---albeit with varying input assumptions. Abundance matching is also used to infer the masses of satellite galaxies by establishing a statistical mapping between a galaxy's stellar mass and the mass of its DM halo~\citep{2018ARA&A..56..435W}, though this technique does not inform the profile shape.

An alternative approach is to perform probabilistic inference of satellite properties using a semi-analytic satellite generator. While typically used to recover statistical properties of the satellite population as a whole~\citep[e.g.,][]{Koposov:2009ru,Li:2009kv,Maccio:2009aek,Font:2011ds,Guo:2010ap,Brooks:2012ah,Starkenburg:2012fh,Barber:2013oua,Pullen:2014gna,Guo:2015jia,Lu:2016geb,DES:2019ltu,Nadler:2022dvo,2023ApJ...948...87W,2022MNRAS.516.3944M,Kravtsov:2023oxa,Ahvazi:2024txq}, they can also be applied to the modeling of individual dwarf galaxies~\citep{Taylor:2000zs,Penarrubia:2010jk,2018PhRvD..97l3002H,Ando:2020yyk,Dekker:2021scf,Horigome:2022gge,Akita:2023yga, Folsom:2023ejk}. Such generators allow one to efficiently populate satellites of a MW-mass halo consistently with the expectations of hierarchical assembly. Typically, semi-analytic models evolve the orbits of these satellites through time, keeping track of tidal mass loss. After generating a large sample of satellites, one can infer unobserved satellite characteristics by conditioning on satellite observables~\citep{Folsom:2023ejk} such as \Mhalf{}, the mass contained within the 3D half-light radius, \rhalf, or \Mstar{}, the stellar mass of the system.  
This procedure yields a probability distribution function~(PDF) from which one can compute quantities of interest using the conditioned ensemble of simulated halos, such as their DM densities. 
In this paper, we use the \SatGen{} satellite generator~\citep{Jiang:2020rdj,Green:2021vkf} to predict the unobserved halo properties of the MW UFDs and the brighter satellites.

\begin{figure}
    \includegraphics[width=0.98\columnwidth]{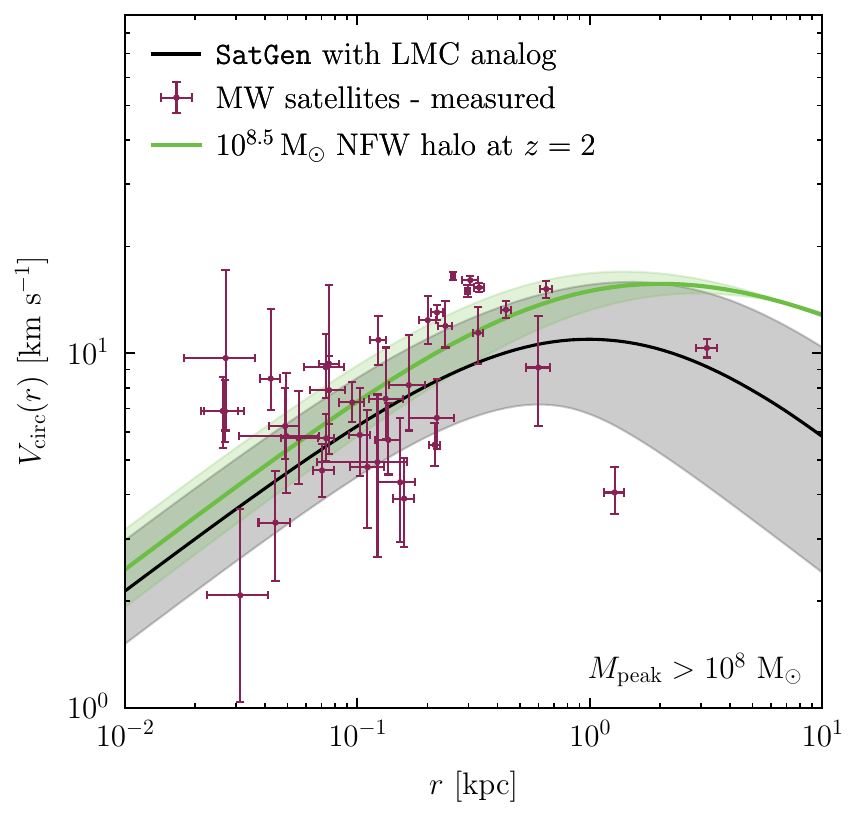}
    \caption{
    The circular velocity \Vcirc{} as a function of radius for the 39 MW satellites considered in this work. 
    The dark magenta crosses reflect the median and 1$\sigma$ bands of \rhalf{} and the estimated \Vcirc(\rhalf) of the observed MW satellite galaxies, computed using the \Wolf{} dynamical mass estimator. 
    For comparison, the \Vcirc{} profile of a median-concentration $\Mvir = 10^{8.5}~\Msun{}$ NFW halo at $z=2$ (a typical infall mass and redshift for a MW UFD) is shown in bright green. The solid line assumes the median of the \Diemer{} \concmass{} relation, and the corresponding band corresponds to the $1\sigma$ scatter of 0.16~dex in this relation. 
    Further, the median~(black line) and 68\% containment band~(gray) of the \SatGen{} satellite $V_\mathrm{circ}$ profiles are shown at each radius for satellite halos whose host has an LMC-analog satellite (as defined in \autoref{sec:model}, though in the bands the LMC-analog satellite itself is removed).  
    The \Vcirc{} values of the observed systems are more diverse than expected from the CDM expectations, with some at low \rhalf{} scattering to larger velocities and others at high \rhalf{} scattering to smaller velocities.
    }
    \label{fig:vcirc}
\end{figure}

One key comparison between observational data and CDM expectations, as set by \SatGen{}, is summarized in \autoref{fig:vcirc}, which compiles the  circular velocities of the known MW dwarfs, estimated at their \rhalf{} from the luminosity-weighted LOS stellar velocity dispersions, $\sigmaLOS$. We take $\Vcirc(\rhalf) = \sqrt{3} \sigmaLOS$, in keeping with the $\Mhalf{}$ estimator of~\citet[][hereafter \Wolf{}]{Wolf:2009tu}. The $\sigmaLOS$ and \rhalf{} values are taken from either the LVDB or KDSA, as detailed in \autoref{sec:data}. As a simple first comparison, we show the $V_{\rm circ}(r)$ profile for a  Navarro--Frenk--White~\citep[NFW;][]{Navarro:1996gj} halo with virial mass $M_{\rm vir}=10^{8.5}~\Msun{}$ at a redshift of $z=2$ in bright green. The median and 68\% containment bands are determined by the concentration--mass~(\concmass{}) relation from~\citet[][hereafter \Diemer{}]{Diemer:2018vmz}  with 0.16~dex scatter. Because of the steep decline of the CDM subhalo mass function, most UFDs should be in similar halos, and this halo mass and infall redshift reflect typical expectations for the MW's UFDs~\citep{Simon_2018, Fillingham:2019rvu, DES:2019ltu, Benitez-Llambay:2020zbo, Nadler:2025mnz}. 

There are two main features of this comparison that are expanded on in this paper. First, while the CDM profiles are expected to exhibit a power-law relation with $V_{\rm circ}(r) \propto r^{0.5}$ at small radii, the data prefer a shallower power-law index.
Second, there are a few UFDs with high central densities that lie close to the top of the predicted uncertainty band or higher, and others that lie below. This observation implies that the inferred density profiles of MW UFDs exhibit a larger variance than expected for CDM.  At this stage, it is not possible to discern whether this is an artifact of large systematic uncertainties in the inferred circular velocities, selection biases against faint, low-dispersion systems, or a genuine mismatch.

Taking this comparison a step further, we show the median and 68\% containment \Vcirc{} profiles for \SatGen{} satellite halos from systems that host an analog of the Large Magellanic Cloud (LMC; as defined in \autoref{sec:model}) in gray. We remove the LMC-analog satellites from these distributions and select for \SatGen{} satellites with an infall virial overdensity mass $\Mpeak{}>10^8~\Msun{}$. The \SatGen{} bands account for the satellite mass function, scatter in the $c-M$ relation, and modifications to the profile shape from tidal mass loss. Importantly, the observations from the previous paragraph hold when comparing the data to the \SatGen{} band. These differences translate into a discrepancy between a dwarf galaxy's halo density  as inferred from its stellar mass as opposed to its stellar kinematics.

This paper is organized as follows. \autoref{sec:methods} describes the satellite model and inference methods, along with the observational data used in this work. \autoref{sec:results} outlines the main results on halo inference, and \autoref{sec:vcirc_rhalf} compares the dwarf galaxy  $V_{\rm circ}-r_{1/2}$ relation to CDM expectations. \autoref{sec:discussion} provides some extended discussion on dwarf discovery prospects and the identification of Ursa Major III/Unions 1.  \autoref{sec:conclusions} concludes. \autoref{app:systematics} details the systematic uncertainties on the inference results, while \autoref{app:tidal_stripping} quantifies tidal mass-loss effects. \autoref{app:lvdb_keckdeimos} compares inference results using LVDB and KDSA data. 
\autoref{app:simulations} compares \SatGen{} halo statistics to two $N$-body simulations, and \autoref{app:powerlaw} analytically derives the power-law behavior of the $V_{\rm circ}-r_{1/2}$ relation. 

\Needspace{4\baselineskip}
\section{Methodology}
\label{sec:methods}

The inference procedure of~\citet{Folsom:2023ejk} predicts properties of satellite halos, such as their density profiles, by leveraging correlations with observables such as stellar kinematics and luminosities. These correlations can be resolved with a large population of semi-analytically generated satellite halos. Heuristically, the density profile of a given MW satellite is inferred by selecting halos from the generated sample that reproduce the observed properties of the dwarf. The density profiles of the synthetic satellites then serve as a model for the actual satellite density. This section describes the semi-analytically generated sample, explains the inference procedure, and reviews the observed data used to select the synthetic satellites.  \autoref{app:systematics} summarizes how varying the assumptions in the semi-analytical model impacts the main results of the study. 

\Needspace{4\baselineskip}
\subsection{Semi-Analytic Satellite Generation}
\label{sec:model}

We generate a large population of synthetic satellite galaxies using 
\SatGen{}~\citep{Jiang:2020rdj,Green:2021vkf},\footnote{\href{https://github.com/JiangFangzhou/SatGen}{https://github.com/JiangFangzhou/\SatGen{}}} which emulates the hierarchical assembly of a host halo (here, of MW mass) via accretion of satellites, each of which is itself assembled from smaller halos. 
The merger trees describing the host halo's growth, along with a number of empirical scaling relations described below, set the initial conditions for the satellite system. \SatGen{} then evolves the satellite population through time under the influence of the host's gravitational potential, which grows self-consistently as the host halo forms. During this orbit integration, \SatGen{} accounts for tidal mass loss and updates the internal structure of the satellites accordingly. This evolution proceeds until the host halo is fully assembled, yielding a realistic population of tidally evolved satellite halos. We run \SatGen{} with the default cosmology assumed in~\citet{Jiang:2020rdj}---a flat $\Lambda$CDM cosmology with $z=0$ matter density $\Omega_m=0.3$, baryon density $\Omega_b=0.045$, dark energy density $\Omega_{\Lambda}=0.7$, power spectrum normalization $\sigma_8=0.8$, power-law spectral index $n_s=1$, and Hubble parameter $h=0.7$.

\begin{deluxetable*}{lcclcc}
\tablecaption{Summary of \SatGen{} suites. \label{tab:simulations}}
\tablehead{
\colhead{Variant} & \colhead{$M_{\rm host}$} & \colhead{$M_{\rm peak,min}$} & \colhead{$c-M$ relation} & \colhead{$N_{\rm hosts}$} & \colhead{$N_{\rm halos}$} \\
\colhead{} & \colhead{$[\Msun{}]$} & \colhead{[\Msun{}]} & \colhead{} & \colhead{} &\colhead{}
}
\startdata
Fiducial & $1 \times 10^{12}$ & $1 \times 10^{8}$ & \Diemer{}, 0.16 dex scatter & $1 \times 10^4$ & $2.4\times 10^6$ \\
$\Mpeak{}>10^7~\Msun{}$ & $1 \times 10^{12}$ & $1 \times 10^{7}$ & \Diemer{}, 0.16 dex scatter & $2 \times 10^3$ & $4.1\times 10^6$ \\
Diemer + 0.12 dex & $1 \times 10^{12}$ & $1 \times 10^{8}$ & \Diemer{}, 0.12 dex scatter & $1 \times 10^4$ & $2.4\times 10^6$ \\
Zhao & $1 \times 10^{12}$ & $1 \times 10^{8}$ & \Zhao{}, intrinsic model scatter & $1 \times 10^4$  & $2.4\times 10^6$ \\
Heavy host & $2 \times 10^{12}$ & $1 \times 10^{8}$ & \Diemer{}, 0.16 dex scatter & $1 \times 10^4$ & $4.8\times 10^6$
\enddata
\tablecomments{Columns are: the suite name; the host halo mass, $M_{\rm host}$; the minimum peak halo mass, $M_{\rm peak,min}$; the choice of $c-M$ relation, with the associated scatter; the number of \SatGen{} host realizations, $N_{\rm hosts}$; the number of \SatGen{} subhalos surviving quality cuts, $N_{\rm halos}$. The last four variants are primarily discussed in \autoref{app:systematics}.}
\end{deluxetable*}

\SatGen{} reproduces the satellite mass functions and circular velocities of hydrodynamical and CDM-only cosmological zoom-in simulations of MW-mass halos~\citep{Jiang:2020rdj,Nadler:2022dvo,Buch:2024ssx}. It has previously been used to study dwarf galaxy quenching~\citep{2023ApJ...949...94G}, the stellar-to-halo mass relation~\citep[SHMR;][]{Monzon:2024rgl,2023ApJ...956....6D}, high-redshift galaxies~\citep{Dekel:2021fpg}, and stellar streams~\citep{Adams:2024zhi,Dropulic:2024tsk}. Compared to cosmological simulations, \SatGen{} is able to rapidly produce large satellite populations; the semi-analytic relations in \SatGen{} act as an effective model for the complicated astrophysics emergent in these simulations. This comes at the cost of modeling assumptions, however. In particular, \SatGen{} describes all halos as spherical, neglects satellite--satellite interactions, and relies on numerical simulations for its calibration. Further, though it is possible to incorporate baryonic feedback prescriptions in the \SatGen{} model, this study is performed with a version of \SatGen{} that uses NFW and tidally truncated NFW halos, without coring from baryonic feedback.  This may affect the DM density profiles of the brighter dwarfs in the sample~($\Mstar \gtrsim 10^6~\Msun{}$). However, for the UFDs, baryonic feedback is unlikely to affect the DM densities~\citep{Fitts:2016usl,Lazar:2020pjs,2025ApJ...995...25M}.

We adopt a $z=0$ host virial mass of $10^{12}~\Msun$, with virial overdensity from~\citet{Bryan:1997dn}.\footnote{All masses in this work are virial overdensity masses defined in this manner, unless otherwise noted.} The halo corresponding to the MW--like host is augmented with a Miyamoto--Nagai disk potential describing the baryonic content of the galaxy~\citep{1975PASJ...27..533M}. The disk's mass grows with the DM halo and modifies the effective concentration of the DM halo as per~\citet{Green:2021vkf}. The disk is 5\% of the halo mass\footnote{At the present day, this is $5\times 10^{10}~\Msun{}$, consistent with the MW's $(4\pm1)\times10^{10}~\Msun$ disk~\citep{Bland-Hawthorn:2016lwg}.} and has a scale radius 12.5 times its scale height; this is the fiducial choice of~\citet{Green:2021vkf} and corresponds to a disk slightly thinner than the MW's thin disk, for which the radius-to-height ratio is 8.7~\citep{Bland-Hawthorn:2016lwg}. Note that~\citet{Green:2021vkf} find that the disk mass is the most important parameter for satellite disruption, with the shape being subdominant. 

The infall times and peak virial masses of the satellites in \SatGen{} are determined by extended Press--Schechter theory~\citep{Lacey:1993fec}, which is calibrated to reproduce the merger trees of $N$-body cosmological simulations~\citep{Parkinson:2007yh}. These merger trees are hierarchical, meaning each satellite will itself assemble from the accretion of smaller halos, a process that iterates until reaching a chosen resolution. In our fiducial run, the resolution limit is taken to be $\Mpeak{}>10^8~\Msun$ (note that a halo's peak mass in \SatGen{} is equivalent to its mass at infall into its original host). This is chosen to reflect the threshold for galaxy formation, but it is possible that lower $\Mpeak{}$ halos could still host galaxies---see, e.g.,~\citet{DES:2019ltu, Benitez-Llambay:2020zbo, Nadler:2025mnz} for further discussion of the galaxy formation threshold. We explore the impact of this threshold in detail in \hyperref[app:mass_floor]{Appendix~\ref*{app:mass_floor}} and comment throughout the main body where results change as this threshold is varied. 

Before infall into their hosts, the satellite galaxies are described by an NFW profile, with concentration parameters randomly drawn from the \Diemer{} \concmass{} relation with a scatter of 0.16~dex. \hyperref[app:concmass]{Appendix~\ref*{app:concmass}} presents the results for other relations, including \Diemer{} with a scatter of 0.12 dex and that of~\citet[][hereafter \Zhao]{Zhao:2008wd}. This choice of scatter is motivated by~\citet{Diemer:2014gba}, who find this scatter in concentration at fixed mass and redshift. The halos are populated with galaxies using a SHMR---see \autoref{sec:inference}.

When a satellite is accreted onto its host, it is given an orbital initial condition calibrated to cosmological simulations~\citep{Li:2020bom}. The orbit is evolved through time under the influence of the host's dynamically evolving gravitational potential, including the influence of dynamical friction~\citep{Chandrasekhar:1943ys}. As the satellites orbit, they are tidally stripped by the host, which preferentially removes mass from the outskirts of the satellite halo. The profiles are updated in response to this mass loss using the transfer function method described in~\citet{Green:2019zkz} and calibrated in~\citet{Green:2021vtc}.\footnote{This calibration differs from~\citet{Du:2024sbt}, who find that tidal stripping does not affect the inner halo as significantly. However, even for fairly extreme mass losses (90\%, cf.~\autoref{fig:tidal_stripping}), the DM density at 150 pc differs by $\lesssim30\%$ between the two models.}
In response to tidal stripping by their host halos, the stellar masses of the satellites are also updated according to the tidal tracks of~\citet{2018MNRAS.481.5073E}. These tidal tracks describe the fraction of stellar mass remaining in the satellite as a function of the halo mass contained within \rmax{} (the radius at which the circular velocity curve reaches its maximum, \vmax{}); they are calibrated to track the baryonic mass loss until 99\% of the initial halo mass within \rmax{} is lost. Satellites only evolve under the influence of their immediate host halo, but they may be released if they are orbiting outside of the host's tidal radius. Consequently, a satellite like the LMC can contribute its own subhalos to the MW satellite population. This mode of accretion is thought to be the origin of some of the MW's UFDs, including Carina~II, Carina~III, Horologium~I, Hydrus~I, Reticulum~II, and Phoenix~II~\citep{2020ApJ...893..121P,2022MNRAS.511.2610C,2022ApJ...940..136P}. 

We perform the following quality cuts on the resulting sample of \SatGen{} satellites:~(i)~select only first-order satellites of the MW host (i.e., we neglect satellites bound to other satellites at redshift $z=0$, but keep those that were contributed by a satellite and have since been released into the MW's halo); (ii)~remove satellites located beyond the virial radius of the MW; (iii)~require that satellites must have retained at least 1\% of their initial halo mass. 

In some cases, we also select for the presence of an LMC-analog satellite by only considering \SatGen{} realizations with a satellite halo of $\Mpeak > 10^{11}~\Msun$ within 100 kpc of the Galactic center. As shown in \hyperref[app:lmc_selection]{Appendix~\ref*{app:lmc_selection}}, selecting for such an LMC-analog satellite affects overall satellite counts, but minimally alters distributions of satellite characteristics and inference results.
Therefore, to maximize the size of the working sample,  we do not restrict to halos with an LMC-analog satellite for the inference procedure described in \autoref{sec:inference}. When comparing MW satellite statistics to \SatGen{}, however, we select for an LMC-analog satellite and note as such.

\autoref{tab:simulations} lists the five unique sets of \SatGen{} runs used in this work, along with the choice of host mass~($M_{\rm host}$), minimum halo mass~($M_{\rm peak,min}$), and \concmass{} relation. The last two columns indicate the number of \SatGen{} hosts simulated, $N_{\rm hosts}$, and the total number of subhalos that pass the quality cuts outlined above, $N_{\rm halos}$. Code used to produce our \SatGen{} runs can be found in our fork of the \SatGen{} GitHub repository.\footnote{\href{https://github.com/folsomde/SatGen}{https://github.com/folsomde/SatGen}}

\needspace{4\baselineskip}
\subsection{Probabilistic Inference Procedure}
\label{sec:inference}

After generating a population of \SatGen{} satellites, we can infer a satellite's DM properties by conditioning on one or more observables of choice. This procedure is reviewed here; we refer the reader to~\citet{Folsom:2023ejk} for more details.  

Given a set of satellite observables, ${\bf \Theta}$, we compute the PDF for unobserved satellite characteristics ${\bf X}$ by marginalizing over $\mathbf{\Theta}$:
\begin{equation}
    f({\bf X}) = \int\! f({\bf X}, \mathbf{\Theta})\ \mathrm{d}\mathbf{\Theta} = \int\! f({\bf X}|\mathbf{\Theta}) f(\mathbf{\Theta})\ \mathrm{d}\mathbf{\Theta} \, ,
\end{equation}
where the distribution $f(\mathbf{\Theta})$ is the PDF inferred from measurements of the observed satellite characteristics. The conditional PDF, $f({\bf X}|\mathbf{\Theta})$, can be estimated using the statistical sample of \SatGen{} galaxies as 
\es{}{
f({\bf X}|\mathbf{\Theta}) \approx f_{\rm pred}({\bf X}|\mathbf{\Theta}) = \frac{f_{\rm pred}({\bf X}, \mathbf{\Theta})}{f_{\rm pred}(\mathbf{\Theta})} \,,
}
where $f_{\rm pred}$ is the distribution of satellites within \SatGen{}. The final PDF for ${\bf X}$ is then estimated by
\begin{equation}\label{eq:inference}
    f({\bf X}) \approx \int\! f_{\rm pred}({\bf X}, \mathbf{\Theta}) \frac{f (\mathbf{\Theta})}{f_{\rm pred}(\mathbf{\Theta})}\ \mathrm{d}\mathbf{\Theta} \,.
\end{equation}
\SatGen{} produces a finite sampling of $f_\mathrm{pred}$, so in practice the integral in~\autoref{eq:inference} is calculated as a weighted average of the sampled values for $\mathbf{X}$, where each \SatGen{} halo is assigned a weight\footnote{As such, we occasionally refer to $f(\mathbf{X})$ as the distribution of $\mathbf{X}$ ``conditioned on $\mathbf{\Theta}$ weights,'' for various $\mathbf{\Theta}$.} $f(\mathbf{\Theta})/f_\mathrm{pred}(\mathbf{\Theta})$ corresponding to how well it reproduces the observed $\mathbf{\Theta}$.  

When considering the distribution of satellites within \SatGen{}, $f_{\rm pred}(\bf \Theta)$, we consider two cases. One approach, denoted ``\Mhalf{} inference'', takes $\mathbf{\Theta} = \log_{10}\Mhalf{}/\Msun$, where \Mhalf{} is the mass contained within the observed half-light radius of the particular dwarf galaxy of interest. We do not rely on \SatGen{}-generated galaxy sizes; instead, we treat an observed $\Mhalf{}$ as a measurement of the enclosed mass of the dwarf halo at the observed half-light radius. Therefore, the \Mhalf{} weights select halos with the same enclosed mass at the observed galaxy's half-light radius. This ensures that the selected satellites reflect the underlying DM halos at a fixed physical radius, rather than also requiring that the \SatGen{} halos have a reasonably-sized galaxy, circumventing the uncertainty in the galactic mass--size relation.

The second approach, denoted ``\Mstar{} inference,'' uses $\mathbf{\Theta} = \log_{10}\Mstar{}/\Msun$, where \Mstar{} is the stellar mass of the dwarf galaxy.  
In particular, the observed \Mstar{} value for a given dwarf galaxy is compared to those of the simulated \SatGen{} halos, which are assigned \Mstar{} values at infall based on a SHMR.
SHMRs considered in the literature vary in their predictions in dwarf galaxy mass ranges ($\Mpeak \lesssim 10^{10}~\Msun$), where even the functional form of the relation is uncertain. We bracket this uncertainty by considering many relations, namely those of~\citet{2018MNRAS.477.1822M,Fattahi:2018ioj, 2023ApJ...956....6D};
and~\citet{2024arXiv240815214K}. In the main body of this work, we restrict consideration to the SHMRs of~\citet[][hereafter \Fattahi{}]{Fattahi:2018ioj} and~\citet[][hereafter \Kim{}]{2024arXiv240815214K}, but results for other relations are provided in \hyperref[app:SHMR]{Appendix~\ref*{app:SHMR}}. The \Fattahi{} relation is based on a simulation of low-mass galaxies ($\Mpeak \lesssim 10^{10}~\Msun$) that predicts stellar masses are exponentially suppressed at low maximum circular velocity, $\vmax$. This choice places even low-luminosity galaxies into fairly massive halos relative to other SHMRs. At the other extreme,~\Kim{} finds a power law that places more dwarf galaxies into smaller halos.

When obtaining the \SatGen{} distribution $f_{\rm pred}({\bf \Theta})$, we produce a continuous PDF by kernel density estimation. Further, for each MW dwarf, we restrict the set of satellites to those with a similar orbital history, as close pericentric passages  result in greater tidal mass loss---see \autoref{app:tidal_stripping} for a quantification of tidal stripping effects and the impact on  satellite inference. Therefore, we restrict to the \SatGen{} subhalos within 50\% to 200\% of the observed dwarf's Galactocentric distance. The results are not highly sensitive to this range. For example, a narrower restriction from 80\% to 125\% shifts the median inferred halo $r_{\rm max}$ and $v_{\rm max}$ by less than 0.01~dex for all MW dwarfs.
The observed distributions for \Mstar{} and \Mhalf{} are taken to be two-sided lognormal distributions centered at the median observed value, with two-sided errors propagated from the observations.

This work considers inference on a range of satellite characteristics $\mathbf{X} \in \{\rmax, \vmax, \rho_{150}, \Mpeak\}$, where $\rho_{150}$ is the DM density at a radius of 150~pc. 

\Needspace{4\baselineskip}
\subsection{Observational Data}
\label{sec:data}

The observed MW satellite census to date comprises 65 confirmed MW dwarf galaxies. Dwarf galaxy summary data (\sigmaLOS{}, \rhalf{}, etc.) are taken from the LVDB version 1.0.5~\citep{Pace:2024sys}.\footnote{\href{https://github.com/apace7/local_volume_database}{https://github.com/apace7/local\_volume\_database}} 
Where available, we use the updated velocity dispersion measurements from the KDSA~\citep{geha2026keckdeimosstellararchivei,geha2026keckdeimosstellararchiveii}---see \autoref{app:lvdb_keckdeimos} for a comparison of results using the two catalogs. 

Since the inference procedure requires a measured LOS velocity dispersion, we restrict this sample to the 43 dwarf galaxies with resolved $\sigmaLOS$ values. We also do not consider the four brightest dwarf galaxies---Fornax,  Sagittarius, the Small Magellanic Cloud~(SMC), and the LMC---in this work. This is because the \SatGen{} sample does not account for baryonic feedback effects, which become increasingly relevant for dwarf galaxies with $\Mstar{} \gtrsim 10^6~\Msun$~\citep{Fitts:2016usl,Lazar:2020pjs,Rey:2023caa,Folsom:2023ejk}. The remaining dwarfs in our sample will likely be less affected, although this is debated, e.g., for Leo~I and Sculptor, the brightest in this set---see~\citet{Strigari:2010un,Amorisco:2011hb,Breddels:2012cq,Richardson:2014mra,Strigari:2014yea,Hayashi:2020jze,Battaglia:2022dii,Pascale:2025zga}.

The final sample therefore contains 39 dwarf galaxies. We split these satellites into three categories: (i)~dwarfs with $\Mstar{}>10^5~\Msun{}$, comprising the classical dwarfs and the ultra-diffuse galaxies Crater~II and Antlia~II~\citep{Torrealba:2016yem,Torrealba:2018fwy}, (ii)~the UFDs, which have $\Mstar < 10^{5}~\Msun$, and (iii)~the satellites (all UFDs) with fewer than 10 member stars with spectroscopic LOS velocity measurements. 
A dwarf's $\Mstar$ is derived from the $V$-band luminosity $L_V$ using a stellar mass--to--light ratio of 1.2~$\Msun/\Lsun$~\citep{Woo:2008gg}. The uncertainty in $\Mstar$ derived from the luminosity is increased by an additional 0.16~dex to capture systematic uncertainties in this conversion~\citep{Woo:2008gg}. Our stellar mass--based definition of UFD yields a population consistent with the recommendation of \citet{Simon:2019nxf}, who defines UFDs as dwarfs with $L_V > 10^5~\Lsun$.

For each MW dwarf, the value for $\Mhalf$ is determined using the \Wolf{} estimator:
\begin{equation}\label{eq:wolf_estimator}
    M_{1/2} \approx 3 G^{-1} \sigmaLOS^2 r_{1/2} \, ,
\end{equation}
where $G$ is the gravitational constant, and $r_{1/2}$ is the 3D deprojected half-light radius, circularized to account for the ellipticity~\citep{Sanders:2016cto}. Specifically, $r_{1/2}=\frac{4}{3}a\sqrt{1-\epsilon}$, where $a$ is the observed semi-major axis of the dwarf galaxy luminosity profile and $\epsilon$ is the ellipticity. The choice of dynamical mass estimator minimally affects the results (see \hyperref[app:mass_estimators]{Appendix~\ref*{app:mass_estimators}}). The circular velocities of the dwarf galaxies in \autoref{fig:vcirc} are computed from $\sigmaLOS$ using the \Mhalf{} estimator: $V_{\rm circ}(r_{1/2})=\sqrt{{G\Mhalf{}}/{r_{1/2}}}=\sqrt{3} \sigmaLOS$.

The velocity dispersions of UFDs in this sample may be subject to systematic
uncertainties arising from membership selection and possible
departures from dynamical equilibrium. Segue~1 is a useful illustration:
earlier studies of the same data reported  different dispersions (within about 20\%) 
\citep{Simon:2010ek, Pace:2018tin, Bonnivard:2015vua}, reflecting
differences in the way membership probabilities were treated. Such differences
should be mitigated by the KDSA analysis, and we adopt its Segue~1 median dispersion value of
4.0~\kms{}, for which \citet{geha2026keckdeimosstellararchivei,
geha2026keckdeimosstellararchiveii} use \emph{Gaia} DR3 proper motions
\citep{2023A&A...674A...1G} to remove foreground stars. This value is close to
the binary-corrected dispersion of 3.7~\kms{} from \citet{Martinez:2010xn},
which is also the value adopted in the LVDB. 

The other effect, departures from
dynamical equilibrium, has also been discussed in the literature as a source
of systematic uncertainty. Willman~1 has an irregular kinematic distribution
that has been read as a sign of tidal disturbance \citep{Willman:2010gy,
Chiu_2026}. 
If Willman~1 (with a half-light radius of about 27 pc) is hosted by a subhalo with mass profile $M(<r)=\Mhalf (r/\rhalf)^2$ (in agreement with \autoref{fig:vcirc}), then we can write the Jacobi (tidal) radius given by $r_J = D\,[M(<r_J)/2M_{\rm MW}(<D)]^{1/3}$~\citep{2008gady.book.....B} at distance $D$ from the MW center as $r_J=\Vcirc^2(\rhalf)D^2/2V_{\rm MW}^2\rhalf$, where we have assumed a flat rotation curve for the MW. The \SatGen{} halos for UFDs have $\Vcirc(30\ \rm pc) > 2 \ \rm km/s$ (see \autoref{fig:vcirc}). Using this with a pericenter distance of 10 kpc (the lower end of the range in \cite{Chiu_2026}, accounting for the LMC) and setting $V_{\rm MW}=200 \ \rm km/s$, we get $r_J = 166$~pc at pericenter, which is significantly larger than the current half-light radius. This shows that it is internally consistent in our analysis to assume that the core of Willman~1 is in dynamical equilibrium today. 
Bo\"otes~III, by contrast,
shows clearer evidence of tidal disruption and association with the Styx
stellar stream \citep{2018ApJ...865....7C}, and \citet{Li:2026ode} revise its
velocity dispersion downward relative to the value adopted here while reporting
an eccentric orbit with pericenter within 10~kpc of the Galactic center. In
the other direction, \citet{2026ApJ...998...47S} find a larger Bo\"otes~I
dispersion than the KDSA from an expanded spectroscopic sample.

Eight of the UFDs considered in this work have spectroscopic measurements for very few stars (e.g., Carina~III with four stars and Tucana~V with three stars). Such small stellar samples may be prone to systematics such as MW contamination.  For this reason, throughout \autoref{sec:results} and \autoref{sec:vcirc_rhalf}, we present results for the low-statistics UFDs separately from those that have more than 10 stars with spectroscopic measurements. 

The full satellite sample and their properties are summarized in \autoref{tab:meta}, which  provides their distance from Earth~($d$), $r_{1/2}$, $\sigmaLOS$, and the system $V$-band luminosity~($L_V$). We also provide quantities inferred from the observed data, including \rmax{}, \vmax{}, and the values of \Mstar{} and \Mhalf{} used to derive these estimates.

\begin{deluxetable*}{lcccccccccc}[!tp]
\tablecaption{Measured and inferred properties of the dwarf galaxies considered in this work.}
\tablewidth{0pt}
\tabletypesize{\small}
\setlength{\tabcolsep}{3pt}
\renewcommand{\arraystretch}{1.0}
\tablehead{
\colhead{Name} & \colhead{$d$} & \colhead{\rhalf{}} & \colhead{\sigmaLOS{}} & \colhead{$\log_{10}\!{L_V}/{\Lsun{}}$} & \colhead{$\log_{10}\!{\Mhalf{}}/{\Msun{}}$} & \colhead{$\log_{10}\!{\Mstar{}}/{\Msun{}}$} & \colhead{$\rmax{} _{,\Mhalf{}}$} & \colhead{$\vmax{} _{,\Mhalf{}}$} & \colhead{$\rmax{} _{,\Mstar{}}$} & \colhead{$\vmax{} _{,\Mstar{}}$} \\
\colhead{} & \colhead{[kpc]} & \colhead{[pc]} & \colhead{[\kms{}]} & \colhead{} & \colhead{} & \colhead{} & \colhead{[kpc]} & \colhead{[\kms{}]} & \colhead{[kpc]} & \colhead{[\kms{}]}
}
\startdata
\multicolumn{11}{c}{$\Mstar{}>10^5~\Msun{}$} \\
\hline
AntII$^b$ & $124^{+5}_{-5}$ & $3200^{+300}_{-300}$ & $6.0^{+0.4}_{-0.4}$ & $5.82^{+0.07}_{-0.07}$ & $7.90^{+0.12}_{-0.12}$ & $5.90^{+0.23}_{-0.23}$ & $1.2^{+0.5}_{-0.3}$ & $12.3^{+1.5}_{-1.2}$ & $2.1^{+1.6}_{-1.1}$ & $24^{+4}_{-8}$ \\
CVnI$^a$ & $211^{+6}_{-6}$ & $436^{+21}_{-21}$ & $7.7^{+0.4}_{-0.4}$ & $5.42^{+0.05}_{-0.05}$ & $7.25^{+0.11}_{-0.11}$ & $5.50^{+0.21}_{-0.21}$ & $1.2^{+1.1}_{-0.5}$ & $16^{+3}_{-2}$ & $2.1^{+1.6}_{-1.1}$ & $23^{+4}_{-7}$ \\
Car$^b$ & $106^{+5}_{-5}$ & $331^{+17}_{-17}$ & $6.6^{+1.2}_{-1.2}$ & $5.70^{+0.06}_{-0.06}$ & $7.00^{+0.19}_{-0.19}$ & $5.78^{+0.22}_{-0.22}$ & $1.1^{+1.0}_{-0.4}$ & $14^{+5}_{-3}$ & $1.9^{+1.5}_{-1.0}$ & $23^{+4}_{-8}$ \\
CraII$^b$ & $117^{+4}_{-4}$ & $1280^{+130}_{-120}$ & $2.3^{+0.4}_{-0.3}$ & $5.21^{+0.07}_{-0.07}$ & $6.69^{+0.19}_{-0.16}$ & $5.28^{+0.23}_{-0.23}$ & $0.36^{+0.14}_{-0.10}$ & $6.1^{+1.1}_{-0.9}$ & $1.8^{+1.4}_{-0.9}$ & $21^{+4}_{-7}$ \\
Dra$^a$ & $81.5^{+1.5}_{-1.5}$ & $258^{+6}_{-6}$ & $9.55^{+0.26}_{-0.25}$ & $5.48^{+0.04}_{-0.04}$ & $7.21^{+0.10}_{-0.10}$ & $5.56^{+0.20}_{-0.20}$ & $1.6^{+2.0}_{-0.8}$ & $24^{+10}_{-5}$ & $1.7^{+1.4}_{-0.9}$ & $21^{+4}_{-8}$ \\
LeoI$^a$ & $258^{+10}_{-10}$ & $306^{+25}_{-25}$ & $9.29^{+0.26}_{-0.26}$ & $6.66^{+0.14}_{-0.14}$ & $7.26^{+0.11}_{-0.11}$ & $6.7^{+0.3}_{-0.3}$ & $1.7^{+2.0}_{-0.8}$ & $22^{+8}_{-4}$ & $3.1^{+2.3}_{-1.4}$ & $29^{+6}_{-7}$ \\
LeoII$^a$ & $233^{+14}_{-14}$ & $220^{+13}_{-13}$ & $7.5^{+0.4}_{-0.4}$ & $5.82^{+0.07}_{-0.07}$ & $6.94^{+0.11}_{-0.11}$ & $5.90^{+0.23}_{-0.23}$ & $1.3^{+1.6}_{-0.6}$ & $19^{+7}_{-3}$ & $2.3^{+1.7}_{-1.1}$ & $24^{+4}_{-7}$ \\
Scl$^a$ & $83.9^{+1.5}_{-1.5}$ & $298^{+6}_{-6}$ & $8.7^{+0.3}_{-0.3}$ & $6.24^{+0.07}_{-0.07}$ & $7.19^{+0.11}_{-0.11}$ & $6.32^{+0.23}_{-0.23}$ & $1.4^{+1.5}_{-0.6}$ & $20^{+7}_{-3}$ & $2.0^{+1.7}_{-1.1}$ & $24^{+5}_{-9}$ \\
Sex$^a$ & $86^{+4}_{-4}$ & $650^{+40}_{-40}$ & $8.8^{+0.5}_{-0.5}$ & $5.42^{+0.06}_{-0.06}$ & $7.54^{+0.11}_{-0.11}$ & $5.50^{+0.22}_{-0.22}$ & $1.3^{+1.0}_{-0.5}$ & $16.8^{+2.8}_{-1.5}$ & $1.7^{+1.4}_{-0.9}$ & $21^{+4}_{-8}$ \\
UMi$^a$ & $70^{+4}_{-4}$ & $334^{+17}_{-17}$ & $8.86^{+0.27}_{-0.24}$ & $5.48^{+0.06}_{-0.06}$ & $7.26^{+0.11}_{-0.11}$ & $5.56^{+0.22}_{-0.22}$ & $1.4^{+1.4}_{-0.6}$ & $20^{+6}_{-3}$ & $1.6^{+1.3}_{-0.8}$ & $21^{+5}_{-8}$ \\
\hline
\multicolumn{11}{c}{Ultra-faints} \\
\hline
Bo\"oI$^a$ & $66.4^{+2.4}_{-2.4}$ & $215^{+11}_{-11}$ & $3.2^{+0.5}_{-0.4}$ & $4.34^{+0.13}_{-0.13}$ & $6.18^{+0.17}_{-0.15}$ & $4.42^{+0.29}_{-0.29}$ & $0.6^{+0.8}_{-0.3}$ & $7.2^{+2.4}_{-1.6}$ & $1.2^{+1.1}_{-0.7}$ & $18^{+4}_{-7}$ \\
Bo\"oII$^a$ & $41.7^{+1.2}_{-1.2}$ & $44^{+7}_{-7}$ & $1.9^{+0.8}_{-0.6}$ & $3.11^{+0.12}_{-0.12}$ & $5.1^{+0.4}_{-0.3}$ & $3.19^{+0.28}_{-0.28}$ & $0.7^{+0.7}_{-0.4}$ & $9^{+5}_{-4}$ & $0.9^{+0.8}_{-0.5}$ & $14^{+4}_{-6}$ \\
Bo\"oIII$^a$ & $46.6^{+0.4}_{-0.4}$ & $600^{+70}_{-70}$ & $5.3^{+2.1}_{-1.7}$ & $4.23^{+0.21}_{-0.21}$ & $7.1^{+0.4}_{-0.3}$ & $4.3^{+0.4}_{-0.4}$ & $0.8^{+0.7}_{-0.4}$ & $10^{+5}_{-3}$ & $1.1^{+1.0}_{-0.6}$ & $16^{+4}_{-7}$ \\
CVnII$^a$ & $160^{+4}_{-4}$ & $73^{+14}_{-14}$ & $5.3^{+1.3}_{-1.0}$ & $4.00^{+0.15}_{-0.15}$ & $6.15^{+0.24}_{-0.21}$ & $4.1^{+0.3}_{-0.3}$ & $1.3^{+1.9}_{-0.6}$ & $20^{+15}_{-6}$ & $1.6^{+1.2}_{-0.8}$ & $19^{+3}_{-6}$ \\
CarII$^b$ & $37.4^{+0.4}_{-0.4}$ & $102^{+10}_{-10}$ & $3.4^{+1.2}_{-0.8}$ & $3.76^{+0.05}_{-0.05}$ & $5.9^{+0.3}_{-0.2}$ & $3.84^{+0.21}_{-0.21}$ & $0.8^{+0.7}_{-0.4}$ & $11^{+6}_{-4}$ & $1.0^{+0.9}_{-0.5}$ & $15^{+4}_{-6}$ \\
CenI$^b$ & $118^{+4}_{-4}$ & $95^{+12}_{-12}$ & $4.2^{+0.6}_{-0.5}$ & $4.07^{+0.10}_{-0.10}$ & $6.07^{+0.17}_{-0.15}$ & $4.15^{+0.26}_{-0.26}$ & $1.1^{+0.9}_{-0.4}$ & $14^{+5}_{-2}$ & $1.5^{+1.2}_{-0.8}$ & $19^{+3}_{-7}$ \\
CB$^a$ & $42.3^{+1.6}_{-1.6}$ & $73^{+6}_{-6}$ & $3.3^{+0.6}_{-0.5}$ & $3.65^{+0.13}_{-0.13}$ & $5.75^{+0.18}_{-0.16}$ & $3.73^{+0.29}_{-0.29}$ & $0.9^{+0.7}_{-0.4}$ & $12^{+4}_{-3}$ & $1.0^{+0.9}_{-0.5}$ & $15^{+4}_{-6}$ \\
EriII$^b$ & $370^{+9}_{-9}$ & $239^{+16}_{-16}$ & $6.9^{+1.2}_{-0.9}$ & $4.78^{+0.14}_{-0.14}$ & $6.90^{+0.18}_{-0.15}$ & $4.86^{+0.30}_{-0.30}$ & $1.3^{+1.5}_{-0.6}$ & $17^{+7}_{-4}$ & $2.1^{+1.5}_{-1.0}$ & $22^{+3}_{-5}$ \\
EriIV$^a$ & $70^{+4}_{-4}$ & $75^{+13}_{-13}$ & $4.5^{+1.1}_{-0.9}$ & $3.35^{+0.14}_{-0.14}$ & $6.04^{+0.25}_{-0.21}$ & $3.43^{+0.30}_{-0.30}$ & $1.1^{+1.2}_{-0.5}$ & $16^{+10}_{-4}$ & $1.2^{+1.0}_{-0.6}$ & $16^{+3}_{-7}$ \\
Her$^a$ & $131^{+6}_{-6}$ & $159^{+17}_{-17}$ & $2.2^{+0.7}_{-0.6}$ & $4.25^{+0.11}_{-0.11}$ & $5.75^{+0.28}_{-0.26}$ & $4.33^{+0.27}_{-0.27}$ & $0.6^{+0.9}_{-0.3}$ & $6^{+4}_{-2}$ & $1.5^{+1.2}_{-0.8}$ & $19^{+3}_{-7}$ \\
HyiI$^b$ & $27.5^{+0.5}_{-0.5}$ & $70^{+9}_{-6}$ & $2.7^{+0.5}_{-0.4}$ & $3.82^{+0.05}_{-0.05}$ & $5.55^{+0.20}_{-0.17}$ & $3.90^{+0.21}_{-0.21}$ & $0.7^{+0.6}_{-0.4}$ & $10^{+3}_{-3}$ & $0.8^{+0.8}_{-0.4}$ & $14^{+5}_{-5}$ \\
LeoIV$^a$ & $151^{+4}_{-4}$ & $136^{+17}_{-17}$ & $3.3^{+0.9}_{-0.7}$ & $3.91^{+0.13}_{-0.13}$ & $6.01^{+0.26}_{-0.21}$ & $3.99^{+0.29}_{-0.29}$ & $1.1^{+0.8}_{-0.6}$ & $10^{+3}_{-3}$ & $1.5^{+1.2}_{-0.8}$ & $19^{+3}_{-6}$ \\
LeoV$^a$ & $169^{+5}_{-5}$ & $49^{+19}_{-19}$ & $3.4^{+1.7}_{-1.0}$ & $3.69^{+0.17}_{-0.17}$ & $5.6^{+0.5}_{-0.3}$ & $3.8^{+0.3}_{-0.3}$ & $1.2^{+1.6}_{-0.6}$ & $16^{+18}_{-5}$ & $1.5^{+1.2}_{-0.8}$ & $18^{+3}_{-6}$ \\
PegIII$^a$ & $215^{+12}_{-12}$ & $110^{+20}_{-17}$ & $2.8^{+1.2}_{-0.9}$ & $3.60^{+0.14}_{-0.12}$ & $5.8^{+0.4}_{-0.3}$ & $3.68^{+0.30}_{-0.28}$ & $1.1^{+0.9}_{-0.6}$ & $10^{+5}_{-4}$ & $1.5^{+1.2}_{-0.8}$ & $18^{+3}_{-6}$ \\
PegIV$^a$ & $89.9^{+1.2}_{-1.2}$ & $56^{+12}_{-9}$ & $3.3^{+1.2}_{-0.9}$ & $3.63^{+0.09}_{-0.09}$ & $5.6^{+0.3}_{-0.3}$ & $3.71^{+0.25}_{-0.25}$ & $1.1^{+1.1}_{-0.5}$ & $14^{+9}_{-4}$ & $1.3^{+1.0}_{-0.7}$ & $17^{+3}_{-7}$ \\
RetII$^b$ & $31.6^{+1.5}_{-1.5}$ & $49^{+7}_{-7}$ & $3.6^{+1.0}_{-0.7}$ & $3.17^{+0.08}_{-0.08}$ & $5.65^{+0.27}_{-0.21}$ & $3.25^{+0.24}_{-0.24}$ & $0.9^{+1.0}_{-0.4}$ & $16^{+9}_{-4}$ & $0.9^{+0.8}_{-0.4}$ & $14^{+5}_{-6}$ \\
Seg1$^a$ & $22.9^{+2.1}_{-2.1}$ & $26^{+4}_{-4}$ & $4.0^{+1.0}_{-0.9}$ & $2.5^{+0.4}_{-0.4}$ & $5.46^{+0.25}_{-0.22}$ & $2.5^{+0.5}_{-0.5}$ & $1.0^{+2.0}_{-0.5}$ & $23^{+22}_{-8}$ & $0.7^{+0.7}_{-0.4}$ & $13^{+4}_{-5}$ \\
TucIV$^b$ & $47^{+4}_{-4}$ & $132^{+24}_{-20}$ & $4.3^{+1.7}_{-1.0}$ & $3.13^{+0.23}_{-0.19}$ & $6.2^{+0.4}_{-0.2}$ & $3.2^{+0.4}_{-0.4}$ & $0.9^{+0.9}_{-0.5}$ & $13^{+9}_{-4}$ & $0.9^{+0.8}_{-0.5}$ & $15^{+4}_{-6}$ \\
UMaI$^a$ & $97^{+6}_{-6}$ & $201^{+16}_{-16}$ & $7.2^{+1.2}_{-1.0}$ & $3.99^{+0.20}_{-0.20}$ & $6.86^{+0.18}_{-0.16}$ & $4.1^{+0.4}_{-0.4}$ & $1.2^{+1.5}_{-0.6}$ & $19^{+9}_{-4}$ & $1.4^{+1.1}_{-0.7}$ & $18^{+3}_{-7}$ \\
UMaII$^a$ & $34.7^{+2.1}_{-2.1}$ & $123^{+10}_{-10}$ & $6.3^{+1.1}_{-0.9}$ & $3.70^{+0.16}_{-0.16}$ & $6.53^{+0.18}_{-0.17}$ & $3.8^{+0.3}_{-0.3}$ & $1.1^{+1.3}_{-0.5}$ & $19^{+10}_{-5}$ & $1.0^{+0.9}_{-0.5}$ & $15^{+4}_{-6}$ \\
Wil1$^a$ & $38^{+7}_{-7}$ & $27^{+6}_{-6}$ & $4.0^{+0.9}_{-0.7}$ & $2.9^{+0.5}_{-0.5}$ & $5.47^{+0.24}_{-0.21}$ & $3.0^{+0.6}_{-0.6}$ & $1.1^{+1.9}_{-0.5}$ & $23^{+20}_{-7}$ & $0.9^{+0.8}_{-0.5}$ & $14^{+4}_{-6}$ \\
\hline
\multicolumn{11}{c}{$N_{\rm stars}<10$} \\
\hline
AqrII$^b$ & $108^{+3}_{-3}$ & $166^{+29}_{-29}$ & $4.7^{+1.8}_{-1.2}$ & $3.68^{+0.08}_{-0.08}$ & $6.4^{+0.4}_{-0.3}$ & $3.76^{+0.24}_{-0.24}$ & $1.1^{+1.1}_{-0.5}$ & $13^{+9}_{-4}$ & $1.3^{+1.1}_{-0.7}$ & $18^{+3}_{-7}$ \\
CarIII$^b$ & $27.8^{+1.3}_{-1.3}$ & $27^{+9}_{-9}$ & $6^{+4}_{-2}$ & $2.89^{+0.12}_{-0.12}$ & $5.8^{+0.7}_{-0.4}$ & $2.97^{+0.28}_{-0.28}$ & $2^{+6}_{-1}$ & $40^{+70}_{-20}$ & $0.8^{+0.7}_{-0.4}$ & $13^{+5}_{-5}$ \\
GruI$^b$ & $126^{+6}_{-6}$ & $150^{+20}_{-30}$ & $2.5^{+1.3}_{-0.8}$ & $3.58^{+0.16}_{-0.16}$ & $5.8^{+0.5}_{-0.3}$ & $3.7^{+0.3}_{-0.3}$ & $0.8^{+0.9}_{-0.4}$ & $8^{+5}_{-4}$ & $1.4^{+1.1}_{-0.7}$ & $18^{+3}_{-6}$ \\
HorI$^b$ & $79^{+4}_{-4}$ & $42^{+4}_{-4}$ & $4.9^{+2.8}_{-0.9}$ & $3.28^{+0.08}_{-0.06}$ & $5.9^{+0.5}_{-0.2}$ & $3.36^{+0.24}_{-0.22}$ & $2^{+4}_{-1}$ & $30^{+40}_{-10}$ & $1.2^{+1.0}_{-0.6}$ & $16^{+3}_{-7}$ \\
LeoVI$^b$ & $111^{+4}_{-6}$ & $120^{+40}_{-60}$ & $2.9^{+1.6}_{-1.3}$ & $3.36^{+0.18}_{-0.24}$ & $5.8^{+0.5}_{-0.5}$ & $3.4^{+0.3}_{-0.4}$ & $0.9^{+1.0}_{-0.5}$ & $9^{+7}_{-5}$ & $1.3^{+1.1}_{-0.7}$ & $17^{+3}_{-6}$ \\
PscII$^b$ & $183^{+14}_{-14}$ & $75^{+8}_{-7}$ & $5^{+4}_{-2}$ & $3.64^{+0.13}_{-0.14}$ & $6.2^{+0.6}_{-0.4}$ & $3.7^{+0.3}_{-0.3}$ & $2^{+3}_{-1}$ & $20^{+40}_{-10}$ & $1.5^{+1.2}_{-0.8}$ & $18^{+3}_{-6}$ \\
TucII$^b$ & $56^{+5}_{-5}$ & $220^{+40}_{-50}$ & $3.8^{+1.1}_{-0.7}$ & $3.43^{+0.12}_{-0.12}$ & $6.34^{+0.28}_{-0.22}$ & $3.51^{+0.28}_{-0.28}$ & $0.8^{+0.7}_{-0.4}$ & $9^{+4}_{-3}$ & $1.0^{+0.9}_{-0.5}$ & $16^{+4}_{-6}$ \\
TucV$^b$ & $55^{+5}_{-5}$ & $31^{+10}_{-9}$ & $1.2^{+0.9}_{-0.6}$ & $2.4^{+0.3}_{-0.3}$ & $4.5^{+0.7}_{-0.5}$ & $2.5^{+0.5}_{-0.4}$ & $0.7^{+0.7}_{-0.3}$ & $7^{+8}_{-4}$ & $0.9^{+0.8}_{-0.5}$ & $14^{+3}_{-6}$ \\
\enddata
\tablecomments{Columns are: galaxy name; heliocentric distance $d$; projected half-light radius \rhalf{}; line-of-sight velocity dispersion \sigmaLOS{}; $V$-band luminosity $\log_{10}(L_V/L_\odot)$; dynamical mass within the half-light radius $\log_{10}(\Mhalf{}/{\rm M}_\odot)$; stellar mass $\log_{10}(\Mstar{}/{\rm M}_\odot)$; and the halo parameters ($\rmax{} _{,\Mhalf{}}$, $\vmax{} _{,\Mhalf{}}$) and ($\rmax{} _{,\Mstar{}}$, $\vmax{} _{,\Mstar{}}$) inferred with the \Mhalf{} and \Mstar{} inference, respectively. Superscripts on galaxy names indicate velocity dispersions from (a) KDSA~\citep{geha2026keckdeimosstellararchivei} or (b) LVDB~\citep{Pace:2024sys}. All other observables are taken from the LVDB. Uncertainties are quoted to one significant figure (two when the leading digit is 1 or 2), with the larger error setting the precision and central values rounded to match. \label{tab:meta}}
\end{deluxetable*}

\Needspace{4\baselineskip}
\section{Dwarf Galaxy Halo Properties}
\label{sec:results}
This section applies the procedure of \autoref{sec:inference} to infer the properties of the MW's observed dwarf galaxies. In particular, we compare results from the kinematic \Mhalf{}-based inference to both the \Mstar{}-inference results and the generic \SatGen{} expectations. For some UFDs, the kinematics prefer DM profiles uncharacteristic of those that host the typical UFD-luminosity galaxy. At a population level, these results suggest that there may be more variance in the densities of MW satellites within 50~kpc of the Galactic center than expected from \SatGen{}. 

\Needspace{4\baselineskip}
\subsection{Structural Parameters}
\label{sec:density}

\begin{figure}[t]
    \includegraphics[width=\columnwidth]{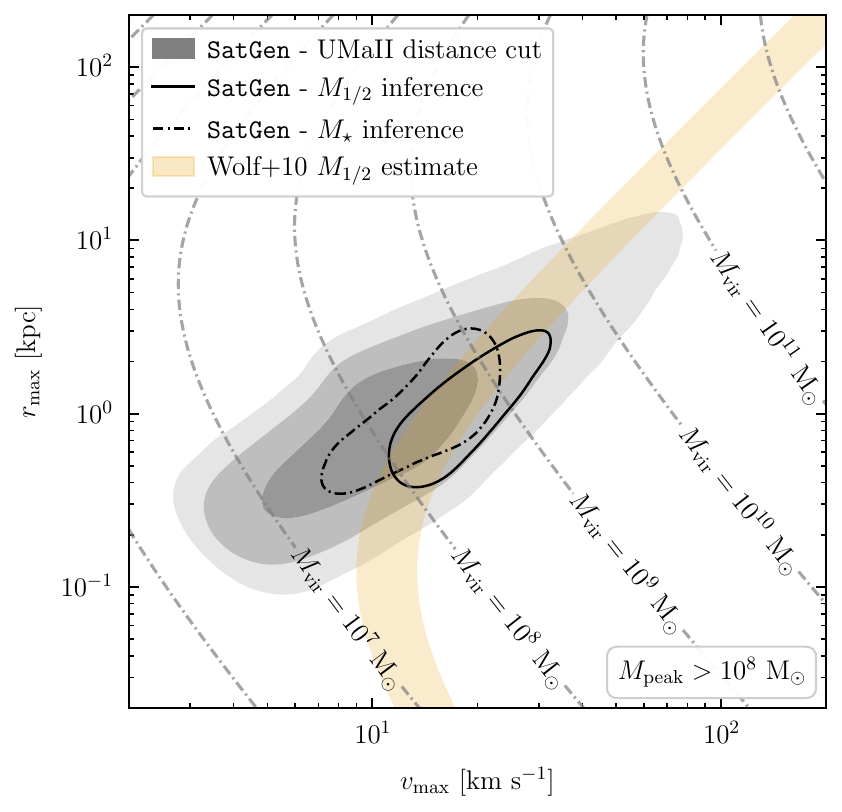}
    \caption{Inferred values for the $z=0$ maximum circular velocity \vmax{} and the radius \rmax{} at which it occurs for the MW dwarf Ursa Major~II~(UMaII). In gray are the 68, 95, and 99.5\% containment regions for the \SatGen{} satellite halos satisfying the distance selection for UMaII.
    The solid and dash-dotted black contours show the 68\% containment for the profile parameters inferred using \Mhalf{} and \Mstar{}, respectively. The \Mstar{} inference assumes the SHMR from \Fattahi{}. Both the \Mhalf{}- and \Mstar{}-inference results are computed using the Fiducial \SatGen{} run and UMaII distance selection. 
    The light orange band shows the locus of NFW halo parameters whose enclosed mass at the half-light radius of UMaII falls within the 68\% confidence interval of the observed $\Mhalf{}$ (as computed using the \Wolf{} estimator), which the \Mhalf{} inference recovers.
    The gray dot-dashed contours denote lines of constant \Mvir{}, the virial mass of a NFW halo with the given parameters at $z=0$.
    }
    \label{fig:rmaxvmax}
\end{figure}

As a concrete example, consider the halo parameters $\mathbf{X} = \{\rmax, \vmax\}$. For an NFW halo, \rmax{} and \vmax{} are sufficient to specify the density profile; while the tidally evolved \SatGen{} profiles are no longer purely NFW, these parameters still provide intuition for the structure of the DM halo. \autoref{fig:rmaxvmax} overlays the inference results at $z=0$ for one UFD, Ursa Major~II, atop the underlying $(\rmax{}, \vmax{})$ distribution for the \SatGen{} halos---from the fiducial $\Mpeak{}>10^8~\Msun{}$ run---satisfying the Ursa Major~II distance cut.   The inferred profile parameters are shown for both \Mhalf{}- and \Mstar{}-inference methods, in solid and dash-dotted black, respectively. For this \Mstar{} inference, we use the \Fattahi{} SHMR. The light orange band shows the $(\rmax{}, \vmax{})$ values for which the corresponding NFW profile has an \Mhalf{}  within the 68\% confidence interval of the observed value. 

\begin{figure*}
    \includegraphics[width=0.98\columnwidth]{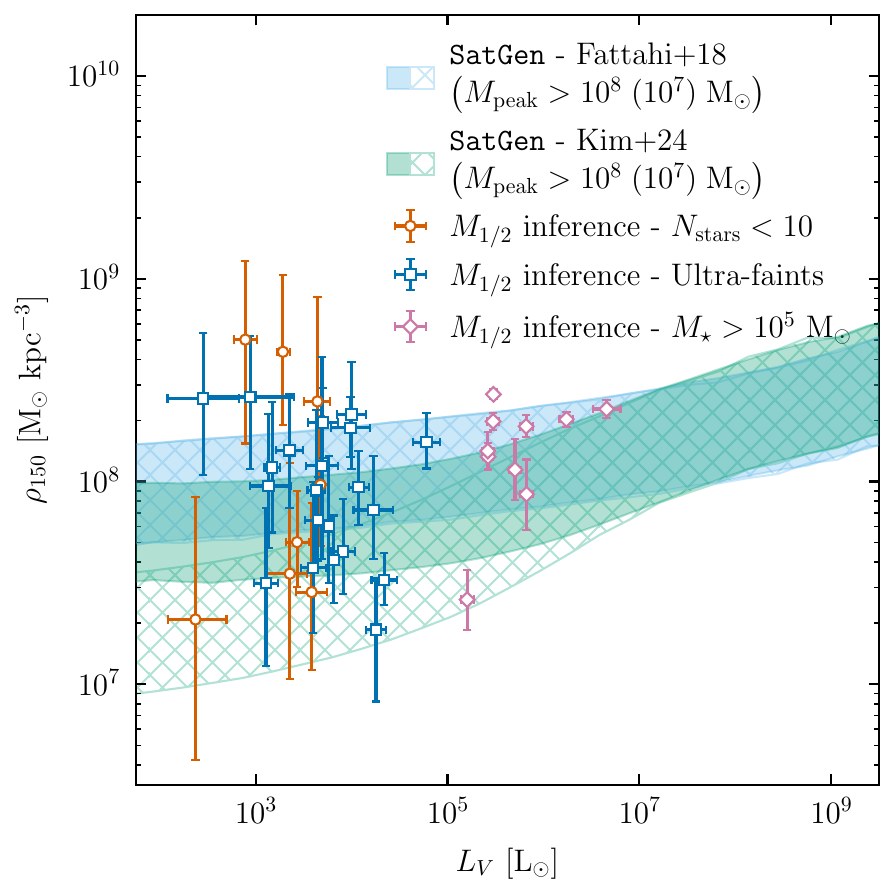}
    \includegraphics[width=0.98\columnwidth]{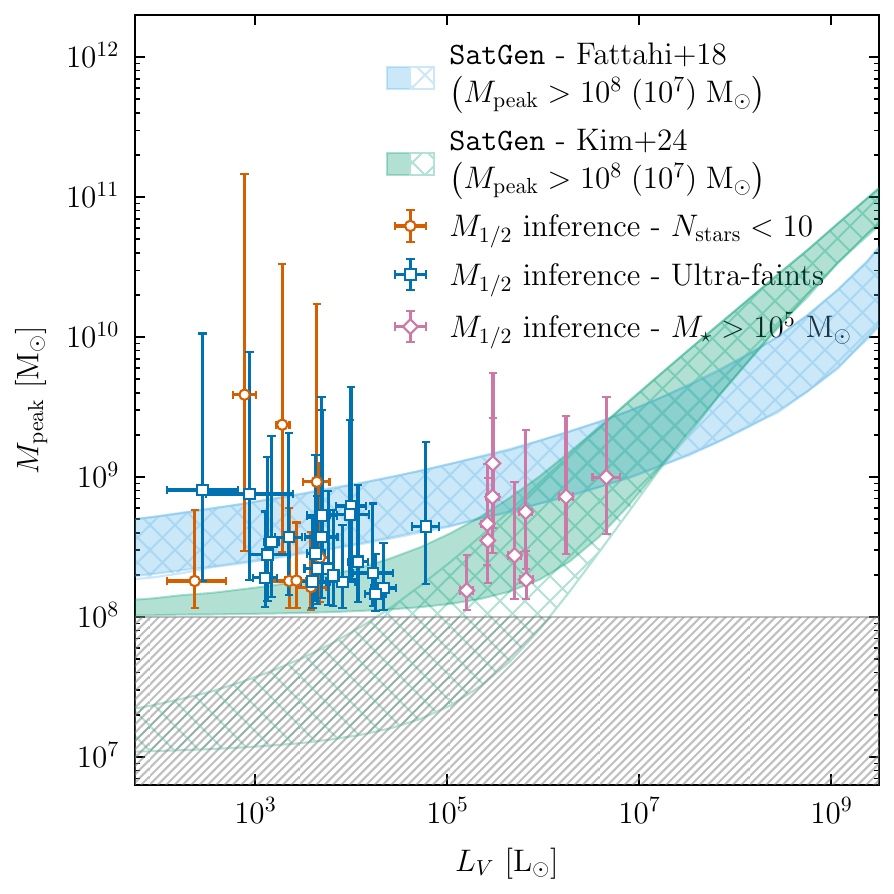}
    \caption{Left:~The observed luminosities and \Mhalf{}-inferred $\rho_{150}$ values for each dwarf galaxy in the sample, split into the three categories: $N_{\rm stars}<10$ (dark orange circles), Ultra-faints (dark blue squares), $\Mstar > 10^5~\Msun$ (magenta diamonds). The inference results are computed using the Fiducial \SatGen{} runs. Overlaid are the 68\% containment bands of \SatGen{} halos populated with stars using the \Fattahi{} and \Kim{} SHMRs in light blue and green, respectively, using the Fiducial \SatGen{} run.  The inferred densities are broadly consistent with the Fiducial SHMR bands, with increasing scatter at lower luminosities. Right:~Same as in the left panel, but for $\Mpeak{}$. In gray, we diagonally hatch the region of $\Mpeak{}<10^8~\Msun{}$ that is not resolved in the Fiducial \SatGen{} run. In both panels, the crosshatching indicates the \SatGen{} bands for the $\Mpeak{}>10^7~\Msun{}$ run. Lowering the \SatGen{} mass floor shifts the \Kim{} $\rho_{150}$ and $\Mpeak{}$ bands to smaller values below $L_V\sim10^7~\Lsun{}$, while the \Fattahi{} bands remain unchanged.  For the inference results based on the $\Mpeak{}>10^7~\Msun{}$ \SatGen{} run, see \hyperref[app:mass_floor]{Appendix~\ref*{app:mass_floor}}.
    }
    \label{fig:L_vs_Mpeak}
\end{figure*}

The \Mhalf{} inference contours for Ursa Major~II are shifted to higher concentrations (smaller \rmax{} for a given \vmax{}) and extend to higher-mass halos (larger \rmax{} and \vmax{}) relative to the \Mstar{} inference. Halos that reproduce the inferred \Mhalf{} are either more concentrated or more massive than those populated by the SHMR with galaxies of the observed stellar mass for Ursa Major~II. While \autoref{fig:rmaxvmax} shows only the \Fattahi{} \Mstar{} inference, the results for the \Kim{} inference are similar. In that case, the \Mstar{}-inference contour retains the same shape as the \Fattahi{} results, but shifts along the central line of the \SatGen{} distribution to lower masses.

Analogues of \autoref{fig:rmaxvmax} for other satellites considered in this work can be found in our GitHub repository.\footnote{ \href{https://github.com/kailashraman/SatelliteDensityInference/}{https://github.com/kailashraman/SatelliteDensityInference/}} In Appendix D of \papertwo{}, we highlight two special cases, Crater~II and Antlia~II, for which the \Mhalf{}- and \Mstar{}-inference results are inconsistent with the Jeans analyses performed in that work.

\Needspace{4\baselineskip}
\subsection{Central Densities and Peak Masses}
\label{sec:rho150}

We now compare the \Mhalf{} and \Mstar{} inferences for all dwarf galaxies. For simplicity, we focus on one-dimensional measures of the DM halo profiles. In particular, the left panel of \autoref{fig:L_vs_Mpeak} shows the \Mhalf{}-inferred values of $\rho_{150}$ within the Fiducial \SatGen{} run for each dwarf galaxy in the sample, plotted against the observed luminosity. The dark orange, dark blue, and magenta symbols refer to the $N_{\rm stars} < 10$, UFD, and $\Mstar{} > 10^5 ~\Msun{}$ samples, respectively.

To compare to the SHMRs, which set the \Mstar{} inference, we populate each \SatGen{} satellite with a stellar mass from the two SHMRs, accounting for tidal stellar mass loss as discussed in \autoref{sec:model}, and illustrate the 68\% containment bands of $\rho_{150}$ as a function of the resulting luminosity. The solid bands correspond to the Fiducial \SatGen{} run. 
The \Fattahi{} relation~(light blue) predicts a sharp cutoff in galaxy formation, with few galaxies in halos with $\Mvir{} \lesssim 10^{8.5}~\Msun$. This preference for larger halos induces larger central densities, so the inferred $\rho_{150}$ values in the ultra-faint regime are larger than what is predicted by the \Kim{} relation~(light green), which continues to host small galaxies in small halos. For context, the 68\% containment for the \Mstar{}-inferred $\rho_{150}$ of a given dwarf would essentially correspond to evaluating this band at the dwarf's observed luminosity, with minor shifts due to the galaxy-specific distance cuts involved in the full \Mstar{} inference.

\begin{figure*}
    \centering
    \includegraphics[width=\textwidth]{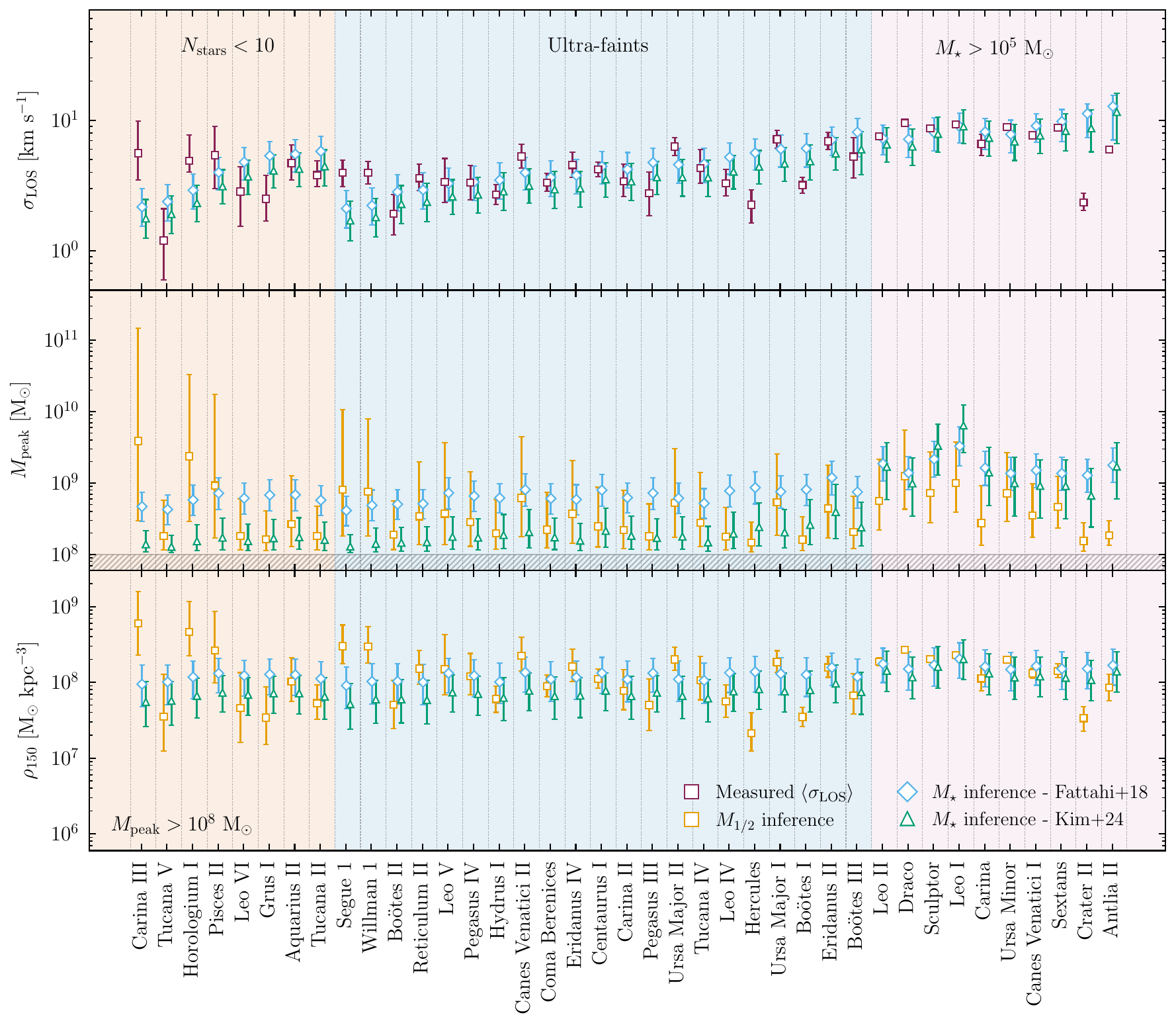}
    \caption{
    The top panel shows the measured velocity dispersion~(dark magenta squares) and the \Fattahi{} and \Kim{} \Mstar{}-inferred velocity dispersions~(light blue diamonds and green triangles, respectively), along with the associated 68\% uncertainties. The MW satellites are ordered by increasing $r_{1/2}$ from left to right within each category ($N_{\rm stars}<10$, Ultra-faints, $\Mstar{} > 10^5~\Msun$).
    The second and third panel are similar but for the inferred \Mpeak{} and the DM density at a radius of 150~pc ($\rho_{150}$), respectively. In these panels, we show the \Mhalf{}-inferred values~(light orange squares) instead of measurements. In the second panel, we also hatch out the region of $\Mpeak{}<10^8~\Msun{}$ in gray.
    For some dwarf galaxies, we see discrepancies between the \Mhalf{}-inferred values and the \Mstar{}-inferred values, with some \Mhalf{}-inferred values scattering upward (Segue 1 and Willman 1) and others scattering downward (Crater II, Hercules, and Bo\"otes I). These discrepancies are all tied to shifts between the measured velocity dispersions and \Mstar{}-inferred dispersions (see main text for details). The inference results in this figure are presented for the Fiducial \SatGen{} run; see \autoref{app:systematics} for a discussion of how results are affected by systematic uncertainties in the semi-analytic modeling.}
    \label{fig:multipanel}
\end{figure*}

For the $\Mstar{} > 10^5~\Msun$ dwarfs~(magenta diamonds), the $\Mhalf{}$-inferred $\rho_{150}$ values are generally consistent with both Fiducial run SHMR bands (although the dwarf galaxies with the smallest $\rho_{150}$ containment intervals lie outside the \Kim{} band). In particular, Crater~II has a much smaller \Mhalf{}-inferred central density ($3\times10^7~\Msun~\text{kpc}^{-3}$) than other galaxies of its luminosity~($1.6\times10^5~\Lsun$). In comparison, there is a much larger spread in the inferred UFD densities~(dark blue squares). Recall that we choose the \Kim{} and \Fattahi{} relations to bracket the systematic uncertainty on the SHMR, so the colored bands together represent the spread in allowed $\rho_{150}$ values within these uncertainties. Even accounting for this range, there are dwarfs for which the inferred \Mhalf{} yields $\rho_{150}$ values outside the predicted range. A few dwarf galaxies (Crater~II, Bo\"otes~I, and Hercules) have small velocity dispersions that lead to inferred densities below both fiducial SHMR bands. There are also two dwarfs with at least 10 stars and LOS velocity measurements, Segue~1 and Willman~1, that lie above the fiducial SHMR bands.

The right panel of \autoref{fig:L_vs_Mpeak} is the same as the left, but for the predicted $\Mpeak{}$. The \Fattahi{} \Mpeak{} band flattens significantly below a luminosity of $\osim 10^7~\Lsun$, reflecting its cutoff in galaxy formation, while the \Kim{} band is cut off by the $10^8~\Msun{}$ mass floor.
The kinematically-inferred masses for the bright dwarfs and ultra-faints typically fall between the two SHMR bands.  Note, however, that the inferred \Mpeak{} for many of the UFDs is close to the $10^8~\Msun$ resolution limit. 

The 68\% containment intervals on $\Mhalf{}$-inferred $\Mpeak{}$ values are quite large. This is because the \Mhalf{} measurement is not particularly correlated with halo mass: a given halo may attain the observed \Mhalf{} by being low--virial mass but concentrated or high--virial mass and diffuse (cf.~\autoref{fig:rmaxvmax}, where the band of profiles consistent with the observed \Mhalf{} spans many orders of magnitude in halo mass). Said another way, while \Mhalf{} provides a mass measurement within the half-light radius, it does not provide much information on the mass contained within the virial radius. Therefore, the \Mhalf{}-inferred halo mass is largely set by the subhalo mass function underpinning the \SatGen{} sample and is in this sense prior-driven. In particular, the subhalo mass function predicts increasingly many low-mass halos, such that the median $\Mpeak$ of the entire satellite sample is near the resolution floor (see \autoref{app:simulations} for more details on the subhalo mass function).

The crosshatched bands in \autoref{fig:L_vs_Mpeak} correspond to the $\Mpeak{}>10^7~\Msun{}$ \SatGen{} runs. This run shifts the \Kim{} bands to lower densities and lower infall masses for $L_V\lesssim10^7~\Lsun{}$ because the \Kim{} SHMR can assign stellar masses greater than  $10^2~\Msun{}$ (i.e., consistent with UFD observations) below $\Mpeak{}=10^8~\Msun{}$. In fact, the \Kim{} power law continues beyond the $10^7~\Msun{}$ mass floor, leading to a flattening of the light green crosshatched band in the right panel at $L_V\lesssim10^4 \ \Lsun{}$. By contrast, the \Fattahi{} bands remain nearly unchanged since the \Fattahi{} SHMR contains an exponential cutoff above the Fiducial \SatGen{} mass floor. Because the choice of mass floor affects the \SatGen{} prior, the $\rho_{150}$ inference results can also shift; however, these shifts do not meaningfully change our conclusions, as discussed in \hyperref[app:mass_floor]{Appendix~\ref*{app:mass_floor}}.
 
\autoref{fig:multipanel} compiles the \sigmaLOS{}~(top panel), $\Mpeak{}$~(second panel), and $\rho_{150}$~(third panel) median values, and associated 68\% uncertainties, for the MW satellites considered in this work. The satellites are separated into the three nominal categories and ordered within those categories by increasing $r_{1/2}$ from left to right. The results are specific to the Fiducial \SatGen{} run. For \sigmaLOS{}, we show the literature values~(dark magenta squares) along with the results from the \Mstar{} inference described in \autoref{sec:inference}~(light blue diamonds and green triangles). Velocity dispersions of \SatGen{} satellites are computed by inverting the \Wolf{} \Mhalf{} estimator. For $\rho_{150}$ and $\Mpeak{}$, the observed values are replaced by results for the \Mhalf{} inference~(light orange squares). Comparing these three panels, one sees that the differences in the $\rho_{150}$ inference broadly match those between the measured and $\Mstar{}$-inferred velocity dispersions, as expected.

For the high-luminosity dwarfs, the \Mhalf{} and \Mstar{} results are generally consistent except for Crater~II.  Of the UFDs with $N_{\rm stars} > 10$, a few stand out for which the \Mhalf{} inference predicts denser halos than the \Mstar{} method. The most prominent upward deviations~($\gtrsim 1\sigma$) occur for Segue~1 and Willman~1.
The \Mhalf{} inference also predicts less dense halos compared to the \Mstar{} method for Hercules and Bo\"otes I, which results from their low velocity dispersions. 

\Needspace{4\baselineskip}
\subsection{Systematic Modeling Uncertainties}

Importantly, the results presented in the previous subsection are robust to systematic modeling uncertainties. As previously mentioned, neither the choice of \concmass{} relation~(\hyperref[app:concmass]{Appendix~\ref*{app:concmass}}) nor SHMR~(\hyperref[app:SHMR]{Appendix~\ref*{app:SHMR}}) affect the conclusions. Changing the choice of dynamical mass estimator~(\hyperref[app:mass_estimators]{Appendix~\ref*{app:mass_estimators}}), modifying the \SatGen{} host mass~(\hyperref[app:host_mass]{Appendix~\ref*{app:host_mass}}), or selecting for \SatGen{} halos from LMC-associated hosts during the inference procedure~(\hyperref[app:lmc_selection]{Appendix~\ref*{app:lmc_selection}}) also minimally change the inferred satellite densities.  

Repeating the analysis on the \SatGen{} $\Mpeak{}>10^7~\Msun{}$ run shifts the \Mhalf{}-inferred $\rho_{150}$ values downward by $\lesssim70\%$ compared to the Fiducial values. The \Mstar{}-inferred values for the \Fattahi{} relation also shift only slightly, moving systematically downward by \mbox{$\lesssim50\%$}, but the results for the \Kim{} relation are more impacted (as expected in keeping with the discussion of \autoref{fig:L_vs_Mpeak}), decreasing by a factor of $\osim2$--3. These shifts tend to maintain or increase the relative differences between \Mhalf{}- and \Mstar{}-inferred values, bringing them into closer alignment in only a limited number of cases which we detail in \autoref{app:mass_floor}; the analysis presented in the previous section is the more conservative choice of \Mpeak{} floor. See \hyperref[app:mass_floor]{Appendix~\ref*{app:mass_floor}} for more details.

To verify that these results are not driven by subhalo mismodeling in the semianalytic framework, we compare the Fiducial \SatGen{} run to the Symphony and Milky Way-est cosmological $N$-body simulations~\citep{Nadler:2022dvo,Buch:2024ssx}, which model  MW-like galaxies down to a subhalo resolution of $\Mvir{}\sim10^8~\Msun{}$.
The \SatGen{} satellite statistics and the distribution of \SatGen{} central densities are comparable to those produced by both of these simulations.
See \autoref{app:simulations} for details.

\Needspace{4\baselineskip}
\section{The $V_{\rm circ}-r_{1/2}$ relation}
\label{sec:vcirc_rhalf}

While the previous section focused on the inferred properties of individual dwarf galaxies, this section looks at the consistency of the dwarf population with CDM predictions.  Specifically, we study the observed dwarf galaxy data at the level of the $V_{\rm circ}-r_{1/2}$ relation and demonstrate a discrepancy between the stellar-kinematics and stellar-mass  predictions. As described previously, we obtain $V_{\rm circ}(r_{1/2})$ from \sigmaLOS{} using the \Wolf{} mass estimator. 

\begin{figure*}
    \centering
    \includegraphics[width=0.49\textwidth]{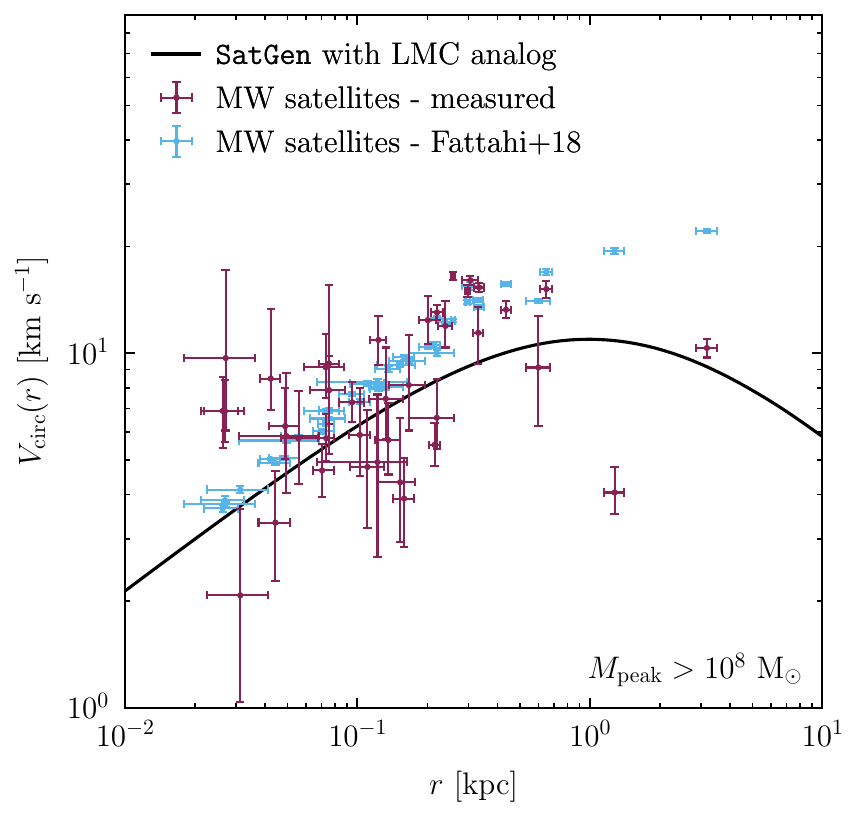}
    \includegraphics[width=0.49\textwidth]{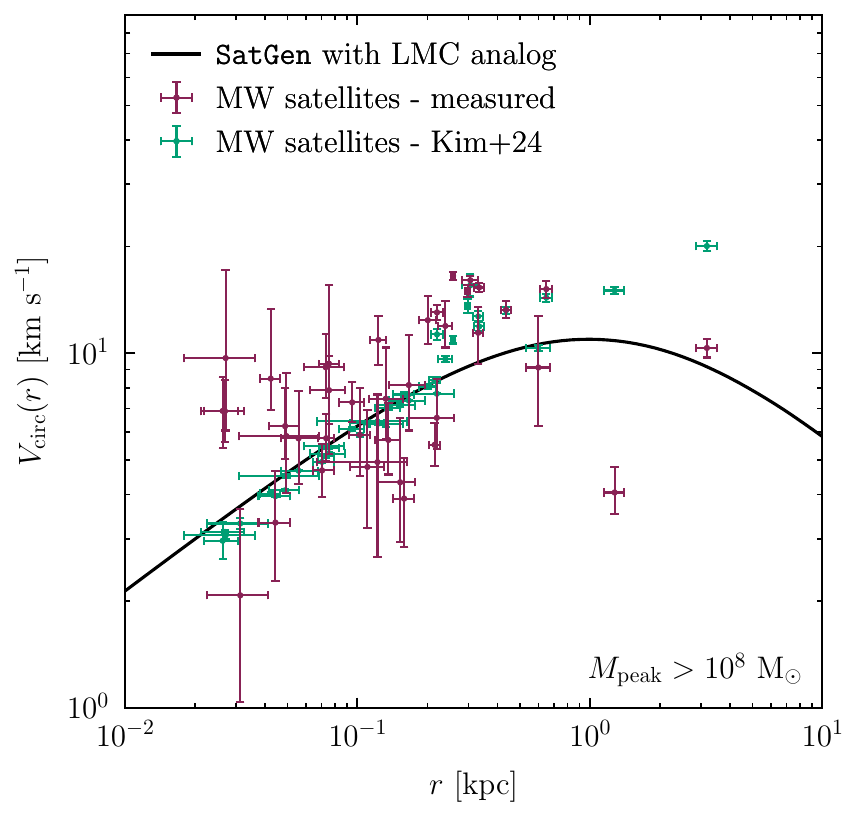}
    \caption{
    Left:  The dark magenta points show the $V_{\rm circ}$ and $r_{1/2}$ values, with measurement uncertainties, for all 39 dwarf galaxies considered in this work.
    The light blue points show the median \Fattahi{}-inferred $V_{\rm circ}$ value and measured $\rhalf{}$, with measurement uncertainties.. The median \SatGen{} $V_{\rm circ}(r)$ profile for the Fiducial \SatGen{} run, selecting for an LMC analog satellite, is shown in black. The \Fattahi{}-inferred results and the \SatGen{} line both follow an approximate power law for $r_{1/2}<300$~pc. Right: Same as the left panel, but showing the \Kim{}-inferred results in light green.
    }
    \label{fig:vcirc_all}
\end{figure*}

\autoref{fig:vcirc_all} plots the $V_{\rm circ}$ and $r_{1/2}$ values for each dwarf galaxy in dark magenta, along with the median Fiducial \SatGen{} $V_{\rm circ}(r)$ line in black (this line matches the median \SatGen{} line in \autoref{fig:vcirc}). The left~(right) panel overlays the median \Fattahi{}-inferred~(\Kim{}-inferred) $V_{\rm circ}$ values in light blue~(green). The error bars shown for the SHMRs reflect the measurement uncertainty in $\rhalf{}$ and the measurement uncertainty on the median \Mstar{}-inferred $V_{\rm circ}$. We do not show the full \Mstar{}-inference posterior width since it reflects the scatter in \SatGen{} halos within a given \Mstar{} interval and not the uncertainty on the median CDM prediction. 

To estimate the measurement uncertainty on the median \Mstar{}-inferred $V_{\rm circ}$ for a given dwarf, we randomly sample from the \Mstar{} PDF extracted from observations and infer a median $V_{\rm circ}$ using the sampled \Mstar{} as a mock-observed value for the stellar mass, keeping the uncertainties in \Mstar{} fixed to their original observed values. We estimate the measurement uncertainty in $V_{\rm circ}$ as the scatter in the median values produced by repeating this procedure $10^2$ times. For \Fattahi{}, all uncertainties are at $\osim0.01$~dex. For \Kim{}, the uncertainty is 0.05~dex for Segue~1, with all other uncertainties below 0.02~dex.  

Within a galactocentric radius of 300~pc, the \SatGen{} line roughly follows a power law, $V_{\rm circ}(r)\propto r^{0.5}$, as expected for an NFW halo within its scale radius. The system of satellites as a whole should also exhibit a power-law relation in these parameters; see \autoref{app:powerlaw} for a derivation of this, along with a prediction for the intrinsic scatter. In general, we expect that any SHMR-inferred power-law will be at least as steep as the NFW relation for the two following reasons. First, galaxies placed in the same-mass halo by the SHMR will collectively follow the NFW power law. Second, more luminous galaxies are typically observed to be of larger size. This results in a positive correlation between $r_{1/2}$ and $V_\mathrm{circ}$, as the more luminous dwarfs will, for any monotonic SHMR, be of larger \Mpeak{} and therefore larger $V_\mathrm{circ}$. The placement of larger-size dwarf galaxies in larger-mass halos steepens the power law thusly.

\begin{figure*}
    \centering
    \includegraphics[width=0.49\textwidth]{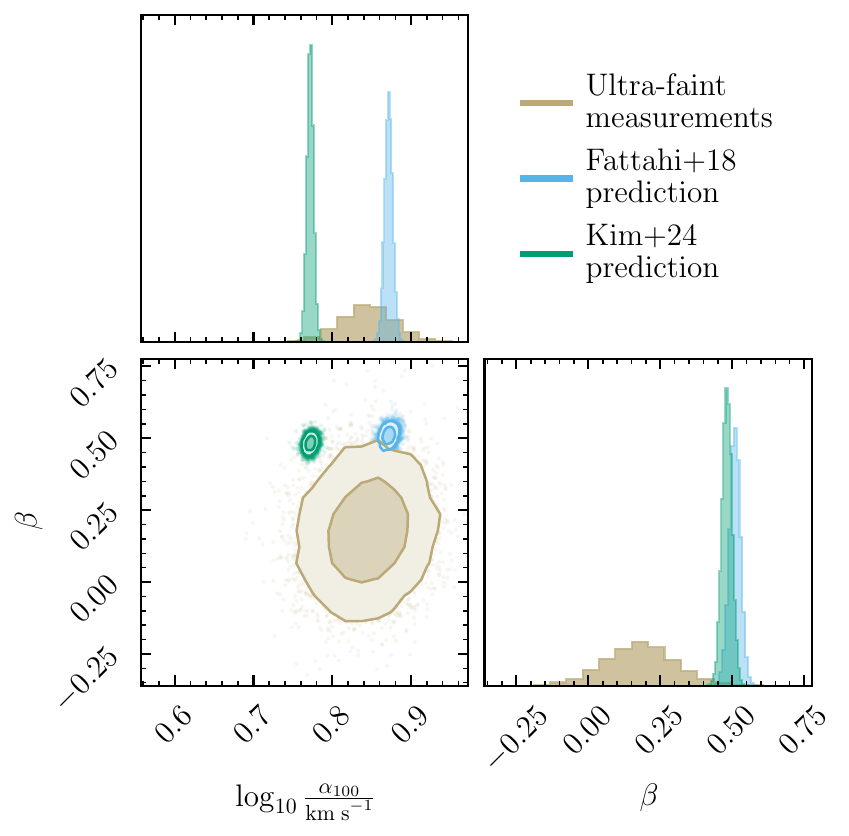}
    \includegraphics[width=0.49\textwidth]{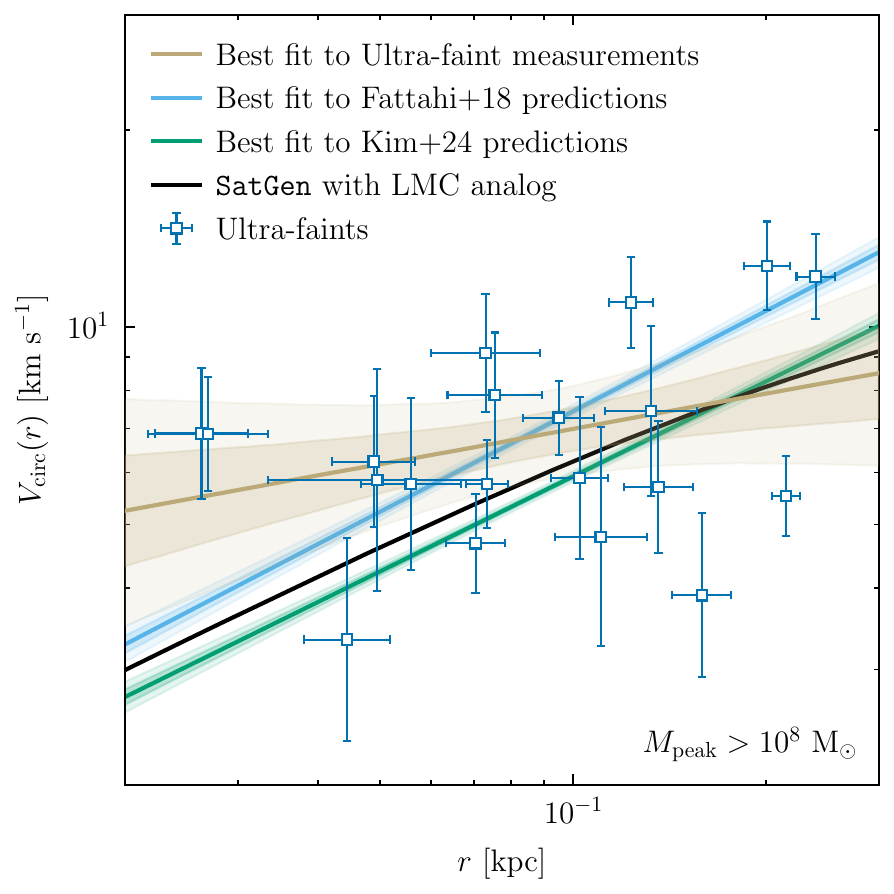}
    \caption{
    Left: Posteriors for the normalization, $\alpha_{100}$, and the power-law index, $\beta$, of the $V_{\rm circ}(r_{1/2})$ measurements for satellites in the Ultra-faints category ($\Mstar{}<10^5~\Msun{}$ and $N_{\rm stars}\geq 10$) with $r_{1/2}<300$~pc, shown in tan. The fit posteriors for the power-law models predicted by \Fattahi{} and \Kim{} are shown in light blue and green, respectively. We additionally show 2D 68\% and 95\% containment contours for each distribution. Right: Power-law model fits to the $V_{\rm circ}(r_{1/2})$ measurements with posterior 68\% and 95\% containment bands~(tan), overlaid on the $V_{\rm circ}$ and $r_{1/2}$ data for those dwarf galaxies in the Ultra-faints category with $r_{1/2}<300$~pc~(dark blue squares). The median and posterior 68\% and 95\% containment bands for the \Fattahi{}~(\Kim{}) inference are shown by the light blue~(green) line/bands, while the median \SatGen{} profile for $\Mpeak{}>10^8~\Msun{}$ is shown in black. The best-fit power-law index for the Ultra-faint measurements is less than the best-fit power-law index for \Fattahi{}~(\Kim{}) at the level of 2.5$\sigma$~(2.4$\sigma$).
    }
    \label{fig:powerlaw}
\end{figure*}

One can test whether the expected $V_{\rm circ}\propto r^{0.5}$ power-law behavior is a good descriptor of the measured $V_{\rm circ}$ values at small $r_{1/2}$~(i.e., dark magenta points). To this end, we model the data as a sample from a log-normal distribution with median value 
\begin{equation}
V_{\rm circ}(r) = \alpha_{100}\left(\frac{r}{100~{\rm pc}}\right)^{\beta} \, ,
\end{equation}
where $\alpha_{100}$ and $\beta$ are free parameters, and intrinsic scatter of $\sigma_{\rm int}$~dex. To ensure reliable data for similar-mass dwarf galaxies within the power-law regime, we restrict to dwarf galaxies in the Ultra-faints category with $r_{1/2}<300~{\rm pc}$ (and $N_{\rm stars} \geq 10$, by construction).

We compute Bayesian posteriors for the three power-law parameters using uniform or log-uniform priors ($0<\log_{10} \frac{\alpha_{100}}{\kms}<2$, $-2<\beta<2$, and $-2<\log_{10} \frac{\sigma_{\rm int}}{\rm dex}<0$). To account for the uncertainties in $r_{1/2}$, we treat the true $r_{1/2}$ values as unknown parameters that are sampled with log-normal priors set by the observed values and uncertainties. The log-likelihood is
\begin{multline}
    \ln \mathcal{L} = -\frac{1}{2} \sum_i \Bigg[ \frac{(\log_{10} V_i - \log_{10}\alpha_{100} - \beta \log_{10} r_i)^2}{\sigma_{V_i}^2 + \sigma_{\rm int}^2} \\
    + \ln \left(2\pi \left[\sigma_{V,i}^2 + \sigma_{\rm int}^2\right]\right) \Bigg] \,.
\end{multline}
Here, $V_i \equiv V_{{\rm circ},i}$ is the circular velocity of the $i$th dwarf galaxy at radius $r_i \equiv r_{1/2,i}$, and $\sigma_{V,i}$ is the corresponding (symmetrized) measurement uncertainty on $\log_{10}\frac{V_i}{\rm km\;s^{-1}}$ in dex.

Before proceeding, we test the assumption of log-normality. The true \SatGen{} $V_{\rm circ}$ distribution at a given radius is a convolution of the log-normal $c-M$ relation with tidal stripping effects that can pull halos to lower $V_{\rm circ}$ at a given radius; therefore, we would expect the true \SatGen{} distribution to have a heavier lower tail than a log-normal distribution. We quantify this effect by fitting a log-normal to the \SatGen{} $V_{\rm circ}(r_{1/2})$ distribution at each dwarf's $r_{1/2}$, with the appropriate distance selection on \SatGen{} halos, and compare the lower-tail probability mass of the fitted log-normal to that of the true \SatGen{} distribution.
Across Ultra-faints with $r_{1/2}<300~\text{pc}$, the probability mass of the fitted log-normal below the 16th~(2.5th) percentile value is at least $98\%$~(83\%) of the true \SatGen{} distribution. Log-normality is therefore a good approximation for the bulk of the satellite distribution, with only mild deviations in the tail. Of relevance for the following discussion, Segue~1 and Willman~1 are log-normal to within $\osim5\%$ past the 2.5th percentile, while Bo\"otes~I and Hercules have $\osim12\%$ deviations from log-normality.

The left panel of \autoref{fig:powerlaw} shows the posteriors on the power-law index, $\beta$, and normalization, $\alpha_{100}$, in tan, generated using the No-U-Turn Sampler~\citep{JMLR:v15:hoffman14a} Markov Chain Monte Carlo algorithm within the \texttt{python} package \texttt{PyMC}~\citep{pymc2023}. The best-fit parameters are $\beta=0.18^{+0.12}_{-0.12}$, $\alpha_{100} = 7.0^{+0.6}_{-0.5}~\kms{}$, 
and $\sigma_{\rm int} = 0.11^{+0.04}_{-0.03}$~dex. This intrinsic scatter (not shown in corner plot) is within the range expected analytically---see \autoref{app:powerlaw}. 

We can repeat this Bayesian analysis for the SHMR-inferred $V_{\rm circ}(r_{1/2})$ values.
The light blue~(green) histograms in \autoref{fig:powerlaw} show the posteriors on $\beta$ and $\alpha_{100}$ for the \Fattahi{}~(\Kim{}) inference. The best-fit power-law indices are $\beta = 0.51^{+0.02}_{-0.02}\;(0.48^{+0.02}_{-0.02})$ for these two models, while the best-fit normalizations are $\alpha_{100}=7.44^{+0.09}_{-0.09}\;(5.92^{+0.06}_{-0.06})~\kms{}$. To compare the posteriors, we compute $P(\beta_{\rm meas}<\beta_{\rm SHMR})$, where $\beta_{\rm meas}$ and $\beta_{\rm SHMR}$ are the best-fit power-law indices for the \sigmaLOS{} measurement and SHMR Bayesian analyses, respectively, and find that $\beta_{\rm meas}<\beta_{\rm Fattahi}\;(\beta_{\rm Kim})$ at the $2.5\sigma$~($2.4\sigma$) level.\footnote{Note that $P(\beta_{\rm meas}<\beta_{\rm SHMR})=P(\beta_{\rm SHMR} -\beta_{\rm meas}>0)$ is computed by sampling a posterior for $\beta_{\rm SHMR} -\beta_{\rm meas}$ from the posteriors for $\beta_{\rm SHMR}$ and $\beta_{\rm meas}$.} The right panel of \autoref{fig:powerlaw} illustrates the power-law fit to the data. The posterior median power law and 68\% and 95\% containment intervals are shown in tan, with the Ultra-faints data overlaid as dark blue squares. The power laws preferred by \Fattahi{} and \Kim{} are also shown in light blue and green, respectively, along with their 68\% and 95\% posterior containment intervals.

The fitted $V_{\rm circ}-r_{1/2}$ relation for the UFDs is sensitive to sample composition; the following analysis explores how the inclusion or exclusion of specific satellites alters the overall results.
First, inclusion of the $\Mstar>10^5~\Msun{}$ dwarf galaxies with $r_{1/2}<300~{\rm pc}$ (Draco, Leo II, and Sculptor), which were initially excluded because this analysis assumes the dwarfs are populated in similar-mass halos, steepens the inferred power-law index to $\beta=0.37^{+0.11}_{-0.11}$. The \Fattahi{} and \Kim{} power-law indices also steepen to $\beta=0.53$ and $\beta=0.54$, respectively.\footnote{Across the analysis variations  considered here, the uncertainty on the \Mstar{} inference power-law indices are $\osim0.02$; we omit the uncertainty on SHMR inference $\beta$ values for the rest of this section for clarity.} In this case, $\beta_{\rm meas}<\beta_{\rm Fattahi}\;(\beta_{\rm Kim})$ at the $1.4\sigma$~($1.5\sigma$) level. Including the $N_{\rm stars}<10$ dwarf galaxies instead leads to a shallower power-law index of $\beta=0.14^{+0.10}_{-0.10}$. The \Fattahi{} and \Kim{} power-law indices are $\beta=0.50$ and $\beta=0.47$, respectively, giving $\beta_{\rm meas}<\beta_{\rm SHMR}$ at the $3.3\sigma$ and $3.2\sigma$ level, respectively. 
Given the high densities inferred for Segue~1 and Willman~1, which are outliers based on our semi-analytic modeling, we also compute results removing these two dwarfs. 
We find a steepening of the inferred power-law index to $\beta=0.31^{+0.16}_{-0.16}$. The \Fattahi{} and \Kim{} power-law indices are $\beta=0.50$ and $\beta=0.48$, respectively. This result gives $\beta_{\rm meas}<\beta_{\rm Fattahi}\;(\beta_{\rm Kim})$ at the 1.2$\sigma$~(1.0$\sigma$) level. 

In addition to the large-dispersion, small--half-light radii dwarfs, the fit to shallow $\beta_\mathrm{meas}$ is in part driven by low-dispersion, large--half-light radii dwarfs (e.g., Hercules and Bo\"otes~I, the two dwarfs closest to the lower-right corner of \autoref{fig:powerlaw}). \citet{geha2026keckdeimosstellararchiveii} explains the discrepancy between these dwarfs' observed dispersions and the CDM expectation through tidal stripping, as they are roughly consistent with the prediction for halos that have lost $\osim99\%$ of their mass~\citep{Penarrubia:2010jk,Esteban:2023xpk}.
Similarly, as shown in
\autoref{app:tidal_stripping}, the observed dispersions of these dwarfs are consistent with halos that have lost $\gtrsim 90\%$ of their mass in the \SatGen{} model, with a preference for halos that are more stripped than the $\osim60\text{--}75\%$ expected mass loss at their galactocentric distances. 
However, our implemented tidal stripping prescription carries increased uncertainty for systems that have low pericenters or have undergone extensive tidal stripping.
Recent work~\citep{Du:2024sbt} argues that \SatGen{} overestimates the reduction of density due to tidal stripping for heavily stripped halos, which would imply that subhalos need to lose even more mass to be consistent with the \Vcirc(\rhalf) values of Hercules and Bo\"otes~I. 
To gauge the impact of Hercules and Bo\"otes~I on the inferred slopes, 
we repeat the analysis without these two dwarfs, finding $\beta\ = 0.33^{+0.11}_{-0.11}$, with $\beta_{\rm meas}<\beta_{\rm Fattahi}\;(\beta_{\rm Kim})$ at the $1.7\sigma$~(1.4$\sigma$) level. 

We also
test how consistent each SHMR is with the data at the level of the 2D $\alpha_{100}$--$\beta$ posteriors. By eye, \Fattahi{} is more
consistent with the data than \Kim{}, given the larger inferred value of
$\alpha_{100}$. We quantify this by computing the Mahalanobis distance~\citep{Mahalanobis1936} of the
zero vector from the two-dimensional distribution of
\begin{equation}
\begin{split}
\Delta X \equiv\ &\left(\log_{10}\alpha_{100},\,\beta\right)_{\rm SHMR} - \left(\log_{10}\alpha_{100},\,\beta\right)_{\rm meas}.
\end{split}
\end{equation}
where, as previously, the subscripts ``meas'' and ``SHMR'' denote results from
the $\sigmaLOS{}$ measurements and the SHMR analyses,
respectively. The Mahalanobis distance is defined as
$d_M \equiv \sqrt{\Delta\mu^{\mathsf{T}} C^{-1} \Delta\mu}$, where
$\Delta\mu \equiv \mu_{\rm SHMR} - \mu_{\rm meas}$ is the difference of the
2D posterior means and $C \equiv C_{\rm SHMR} + C_{\rm meas}$ is the sum of the two
posterior covariance matrices. 

Under the assumption that the two
$\left(\log_{10}\tfrac{\alpha_{100}}{\rm km\;s^{-1}},\,\beta\right)$ posteriors
are independent multivariate Gaussians and the null hypothesis that the true difference between the SHMR prediction and the measurement is the zero vector, $d_M^2$ is $\chi^2$-distributed with two degrees of freedom. We can then quantify consistency by computing $P\!\left(\chi^2_2 \geq d_M^2\right)$ for each SHMR. For the fiducial case including only the Ultra-faints category, \Fattahi{} is inconsistent with the measurements at the $2.2\sigma$ level, while \Kim{} is inconsistent with the measurements at the $3.0\sigma$ level. We conduct this test for the additional SHMRs considered in \autoref{app:SHMR} and find that \Fattahi{} is the most consistent with the velocity dispersion data out of every SHMR considered.

Lastly, we emphasize that these fits are computed for an observationally incomplete sample of dwarf galaxies. In addition to the discoveries of next-generation surveys, there exist several systems with spectroscopic measurements of greater than 10 stars but unresolved velocity dispersions (e.g., Draco~II and Segue~2), which are not used in this analysis. The non-inclusion of such satellites may affect the results presented in this section.
As unbiased and resolved velocity dispersions are collected for more satellites, this procedure will become a more robust method for testing observations against CDM expectations.

\Needspace{4\baselineskip}
\section{Discussion}
\label{sec:discussion}

This discussion focuses on two additional applications of the semi-analytic modeling procedure introduced in this work.  The first (\autoref{sec:dwarfdiscovery}) provides a means of testing future discovery prospects for UFDs.  The second (\autoref{sec:ufcs}) is a way to verify whether a newly discovered system is consistent with CDM expectations, using Ursa Major III/Unions 1 as an esample.

\subsection{Dwarf Discovery Prospects}
\label{sec:dwarfdiscovery}

\begin{figure}
    \includegraphics[width=0.98\columnwidth]{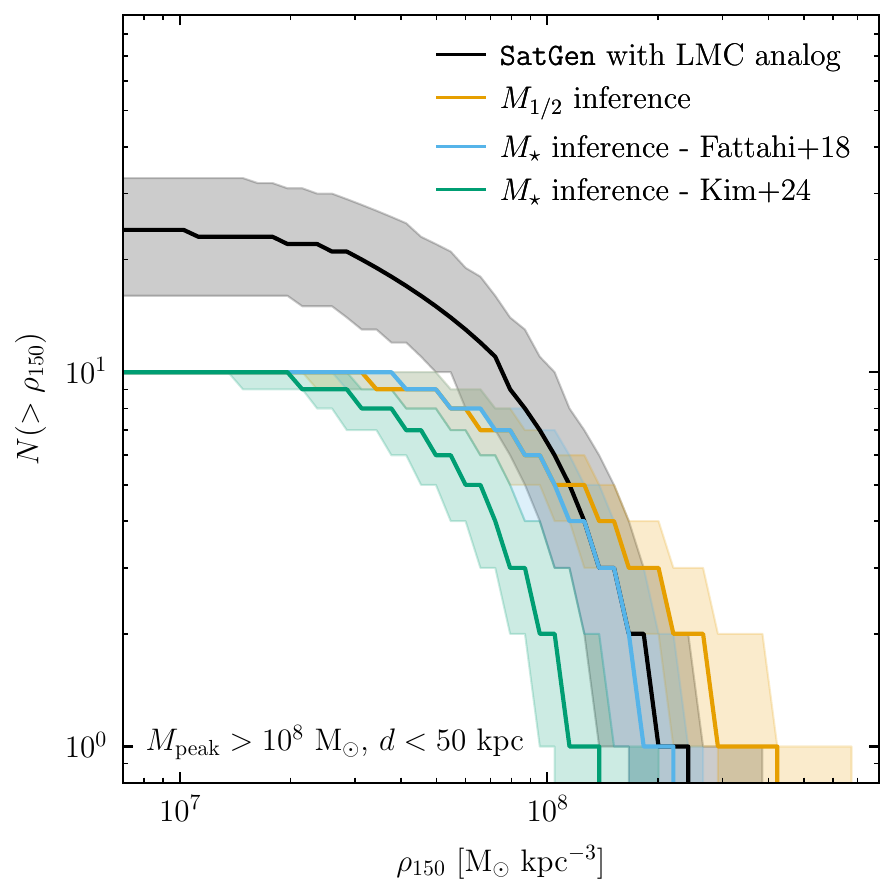}
    \caption{
    The median number of satellites within a galactocentric distance of 50 kpc above a given $\rho_{150}$ threshold, along with 68\% containment bands. The results for the Fiducial \SatGen{} model, selecting for LMC-associated hosts, are shown in gray, as well as  the MW dwarf galaxies conditioned on \Mhalf{} weights~(light orange) and \Mstar{} weights from the \Fattahi{}~(light blue) and \Kim{}~(light green) SHMRs. Results are only shown for MW satellites with $N_{\rm stars} > 10$. The \SatGen{} results are not corrected for observational completeness. The estimated dwarf galaxy central densities conditioning on dwarf galaxy kinematics (\Mhalf{} weights) are higher than both the \SatGen{} expectation and the results conditioning on stellar masses.  The \Mhalf{}-inferred results saturate the high-density \SatGen{} expectation, indicating that we have likely already found the densest MW satellites within 50~kpc, if they are accurate.  There may be more room for discovery if one assumes the \Kim{} results to be true.
    }
    \label{fig:population}
\end{figure}

\autoref{fig:population} shows the cumulative distribution of the $\rho_{150}$ values for the Fiducial \SatGen{} run. 
This figure focuses specifically on the 10 MW dwarfs within 50~kpc of the Galactic center with $N_{\rm stars} > 10$, all of which happen to be ultra-faint. (As discussed in \autoref{sec:data}, we neglect Sagittarius and the LMC.) For each dwarf, we sample a value from its inferred $\rho_{150}$ PDF. These 10 $\rho_{150}$ values are then summarized as a cumulative distribution. We repeat the procedure $10^3$ times and show the median and 68\% containment interval of these distributions in the figure. The band produced using \Mhalf{}-inferred $\rho_{150}$ values is shown in light orange, and the corresponding results for the \Mstar{} inference assuming the \Fattahi{} and \Kim{} relations are shown in light blue and green, respectively.  The figure also shows the distribution for all \SatGen{} hosts that contain an LMC analog in gray, with those LMC analogs removed from the distribution.  Importantly, the \SatGen{} sample is not corrected for observational completeness, so care must be taken in comparing the results with those based on the actual MW dwarfs.  While the 50~kpc distance cut facilitates a more equitable comparison, the \SatGen{} prediction should be treated as an overestimate.  

Across \SatGen{} hosts with an LMC analog, there are $24^{+9}_{-8}$ satellites within a galactocentric distance of 50~kpc. The \SatGen{} median drops to zero at $\rho_{150}\sim2.6\times 10^8~\Msun{}~{\rm kpc}^{-3}$.
In contrast, the \Kim{} inference for the observed satellites drops to zero at $\rho_{150}\sim1.5\times 10^8~\Msun{}~{\rm kpc}^{-3}$, while the \Fattahi{} inference falls off at roughly the same point as the \SatGen{} median. The \Mhalf{} inference finds $\osim2$ satellites with densities above the \SatGen{} cutoff. The two satellites supporting the \Mhalf{} band at these high densities are Segue~1 and Willman~1.  Given this mild overabundance of high-density MW satellites relative to \SatGen{}, we expect to have already seen the densest MW satellites within 50~kpc of the Galactic center if the $\Mhalf$-inferred densities are accurate. If, instead, the \Kim{} results are more realistic, then there may be more to discover in this region, as the observed dwarfs do not saturate the \SatGen{} expectation.

\Needspace{4\baselineskip}
\subsection{Ursa Major~III/Unions~1}
\label{sec:ufcs}

In addition to allowing inference of unobserved dwarf galaxy properties from measured observables, \SatGen{} provides a testbed for evaluating whether measured stellar system observables are consistent with CDM expectations for dwarf galaxies. As a concrete test case, this subsection focuses on the velocity dispersion and $r_{1/2}$ of one ultra-faint compact stellar system, Ursa Major~III/Unions~1~(UMa3/U1). 

\begin{figure}
    \includegraphics[width=0.98\columnwidth]{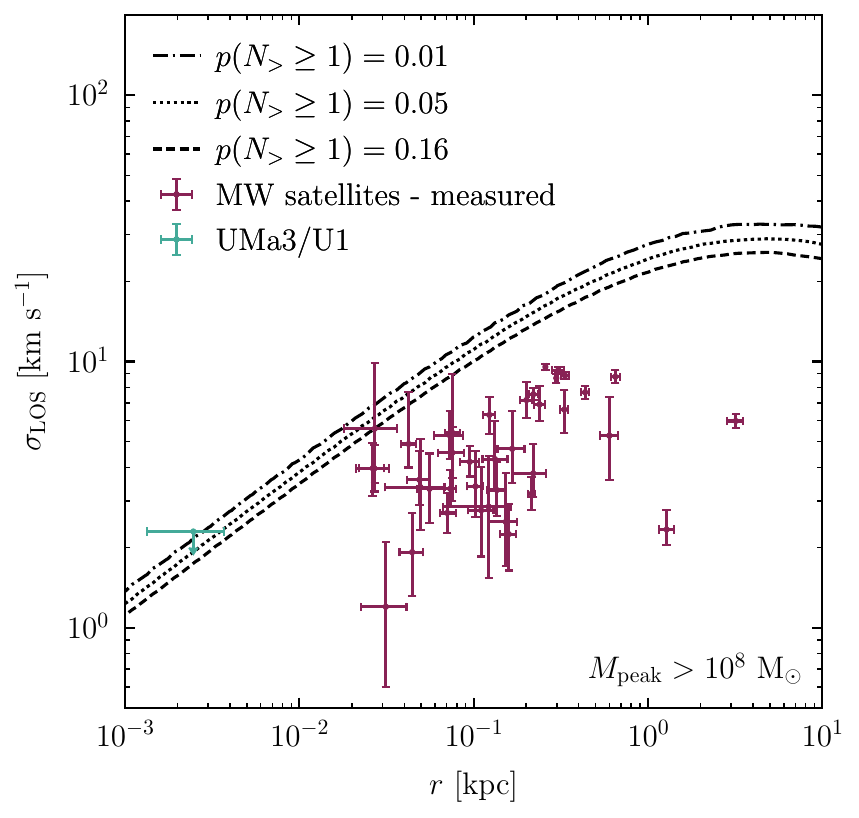}
    \caption{Dwarf galaxy velocity dispersions and half-light radii, along with various \SatGen{} upper limits on satellite statistics.
    The dark magenta points show the velocity dispersions and half-light radii, with measurement uncertainties, for the 39 MW satellite galaxies considered in this work. The velocity dispersion and half-light radius for UMa3/U1~\citep{2025arXiv251002431C} are shown as a 95\% upper limit and 68\% containment interval, respectively, in teal. We compare to \SatGen{} expectations (for the Fiducial run) by computing upper limits on the number of satellite halos with velocity dispersions above a given value at a given halo radius $r$. In the legend, $N_>$ indicates the number of satellites with \sigmaLOS{} above a given value at a given $r$.
    We find that 16, 5, and 1\% of LMC-associated \SatGen{} hosts have at least one satellite halo with a velocity dispersion above the dashed, dotted, and dot-dashed lines, respectively.
    }
    \label{fig:dispersion_limits}
\end{figure}

UMa3/U1 has multi-epoch spectroscopic measurements of LOS velocities~\citep{2024ApJ...961...92S,2025arXiv251002431C}. The original discovery work on UMa3/U1~\citep{2024ApJ...961...92S} found a resolved LOS velocity dispersion of $3.7^{+1.4}_{-1.0}$~\kms{} and $r_{1/2}$ of $\osim3\pm1$ pc, leading to speculation that UMa3/U1 is a compact remnant of a dwarf galaxy hosted by an ultra-dense DM halo~\citep{Crnogorcevic:2023ijs,Errani:2023sgd}. However,~\citet{2025arXiv251002431C} confirmed one UMa3/U1 star as a binary system with a second epoch of spectroscopic measurements and revised the velocity dispersion measurement to only set an upper bound, $\sigmaLOS{} < 2.3$~\kms{} at 95\% confidence. They additionally found a small metallicity spread and concluded that UMa3/U1 is consistent with being a stellar cluster not hosted by a DM halo.

Here, we test the likelihood of producing a \SatGen{} satellite with a velocity dispersion as large as the current UMa3/U1 upper limit. \autoref{fig:dispersion_limits} shows the measured velocity dispersions and half-light radii of MW satellites~(dark magenta), with the dispersion for UMa3/U1 shown as an upper limit~(teal). The lines correspond to various upper limits on the number of satellites with a velocity dispersion above a given value at a given radius, for Fiducial run \SatGen{} hosts with an LMC.  In particular, the dashed, dotted, and dot-dashed lines correspond to a probability of 16, 5, and 1\%, respectively, of finding a \SatGen{} satellite of given dispersion and size. 

The UMa3/U1 upper limit is above the dotted curve at all $r_{1/2}$ values within the measured 68\% containment interval, indicating that saturating the UMa3/U1 velocity dispersion upper limit would be unlikely at the 5\% level according to \SatGen{}. 
We conclude that even if UMa3/U1 is indeed hosted by a DM halo, it is highly unlikely to have such a large velocity dispersion.

\Needspace{4\baselineskip}
\section{Conclusions}
\label{sec:conclusions}

Modeling the DM distribution of dwarf galaxies, particularly in the ultra-faint regime, is critical for testing CDM. In this work, we used the \SatGen{} semi-analytic satellite generator~\citep{Jiang:2020rdj,Green:2021vkf} to infer the DM halo properties of 39 MW dwarf galaxies. Following the framework of~\citet{Folsom:2023ejk}, we conditioned the generated satellites on observed properties to obtain PDFs for DM halo parameters, including the central density~($\rho_{150}$) and peak halo mass~($\Mpeak{}$).

We performed this inference using two distinct weighting criteria. The \Mhalf{} inference conditions \SatGen{} halos to match the dwarf's mass enclosed within the 3D half-light radius, derived from the LOS stellar velocity dispersion via the \Wolf{} estimator (note that we check other dynamical mass estimators in \autoref{app:mass_estimators}). The \Mstar{} inference instead takes \SatGen{} distributions with stellar masses drawn from a particular SHMR and finds the halos that host galaxies of the observed \Mstar{}. To bracket the systematic uncertainty on the SHMR, we present the primary results using the relation from \Fattahi{}, which predicts a sharp galaxy formation cutoff at $\Mvir \lesssim 10^{8.5}~\Msun{}$, and contrast to the \Kim{} relation, which predicts galaxy formation down to lower-mass halos.

The $M_{1/2}$ and $M_*$ inferences yield qualitatively different distributions of UFD central densities. The $M_{1/2}$ inference produces a larger spread in $\rho_{150}$ than expected from the $M_*$ inference. Specifically, the most compact UFDs (e.g., Segue~1, Willman~1, Carina~III, Horologium~I) prefer halos denser than the SHMR-conditioned expectation, while the most diffuse systems (e.g., Crater~II, Hercules, Boötes~I) prefer halos less dense than expected (see \autoref{fig:L_vs_Mpeak} and \autoref{fig:multipanel}). 

At the population level, this variance translates into a relatively flat $V_{\mathrm{circ}} - r_{1/2}$ median relation among the ultra-faints with $r_{1/2} < 300\mathrm{\;pc}$ (see \autoref{fig:powerlaw}). In contrast, the CDM expectation based on \SatGen, and also derived analytically in \autoref{app:powerlaw}, is approximately $V_{\mathrm{circ}}(\rhalf) \propto (
\rhalf)^{0.5}$ with an intrinsic scatter of 0.1--0.15 dex due to the scatter in the concentrations, infall redshifts, and host halo masses. 
This prediction reflects the underlying density profile scaling as $1/r$ and the lack of strong dependence on redshift or halo mass for the \Vcirc{} of UFD host halos at small radii. Notably, the observed power-law index is in $\osim2.4\sigma$ tension with this theoretical expectation.

The discrepancy in central densities remains robust against the systematic variations explored in \autoref{app:systematics}, including the choice of \concmass{} relation, SHMR, dynamical mass estimator, host halo mass, LMC selection, and \SatGen{} infall mass threshold. 
Additionally, as shown in \autoref{app:simulations}, $N$-body simulations such as Symphony~\citep{Nadler:2022dvo} and Milky Way-est~\citep{Buch:2024ssx} produce similar satellite statistics and central densities to \SatGen{}.

However, three key systematics remain outside the scope of our formalism. First, the current sample of dwarfs is  observationally incomplete, which can bias the conclusions regarding the population-level properties of the MW's satellites. Second, kinematically inferred densities may be biased by overestimated velocity dispersions in certain UFDs due to binary contamination, low-statistics biases, or uncalibrated measurement uncertainties. Recent work~\citep{arroyopolonio2026estimatingdynamicalmassesdwarf} suggests that binary contamination can inflate dispersions by 70–80\% in UFDs with $\osim10$--20 spectroscopic measurements, though this effect is mitigated in systems with multi-epoch data, such as Segue~1 and Reticulum~II. Third, our framework is limited by the assumptions inherent to the semi-analytical modeling in \SatGen{}. For example, \SatGen{} assumes all halos are spherical, neglects satellite--satellite interactions, relies on simulation-calibrated tidal tracks, and omits dynamical effects such as the LMC wake~\citep[e.g.,][]{2019ApJ...884...51G}.  

UFDs serve as an important testbed for evaluating the viability of CDM and its alternatives. As more dwarfs are discovered and the inferences of the velocity dispersions of each dwarf improve, the ensemble of UFDs will be a critical test of DM physics. The semi-analytical framework presented in this work and the new procedure to compare an ensemble of mass measurements to theory expectations provide a robust method
for determining the consistency of kinematically derived properties of these dwarfs with CDM. As next-generation surveys yield more precise observational data, this framework will enable increasingly incisive tests of DM models at the smallest galactic scales. 

\Needspace{4\baselineskip}
\section*{\textbf{Acknowledgments}}
We thank Yujin Park for collaboration during the early stages of this project, as well as Marla Geha, Andrew Pace, Nicholas Rodd, Daniel Hooper, and Alex Drlica-Wagner for useful discussions and comments. K.R. and B.R.S. are supported in part by the DOE award
DESC0025293. B.R.S. acknowledges support from the
Alfred P. Sloan Foundation and hospitality from the CERN theory group, where much of this work was completed.  M.L. and D.F. are supported by the Department of Energy~(DOE) under Award No.~DE-SC0007968. M.L. is also supported by the Simons Investigator in Physics Award. M.K is supported by the National Science Foundation award PHY-2514888. D.F. is additionally supported by the Joseph H. Taylor Graduate Student Fellowship. This research used resources of the National Energy Research
Scientific Computing Center (NERSC), a U.S. Department of Energy Office of Science User Facility located
at Lawrence Berkeley National Laboratory, operated under Contract No.~DE-AC02-05CH11231 using NERSC
award HEP-ERCAP0023978. This research has made use of the Keck Observatory Archive (KOA), which is operated by the W. M. Keck Observatory and the NASA Exoplanet Science Institute (NExScI), under contract with the National Aeronautics and Space Administration.

This report was prepared as an account of work sponsored by an agency of the United States Government. Neither the United States Government nor any agency thereof, nor any of their employees, makes any warranty, express or implied, or assumes any legal liability or responsibility for the accuracy, completeness, or usefulness of any information, apparatus, product, or process disclosed, or represents that its use would not infringe privately owned rights. Reference herein to any specific commercial product, process, or service by trade name, trademark, manufacturer, or otherwise does not necessarily constitute or imply its endorsement, recommendation, or favoring by the United States Government or any agency thereof. The views and opinions of authors expressed herein do not necessarily state or reflect those of the United States Government or any agency thereof.

\Needspace{4\baselineskip}
\bibliographystyle{aasjournal}
\bibliography{refs.bib}

\clearpage
\appendix
\renewcommand\thefigure{\thesection\arabic{figure}} 

\Needspace{4\baselineskip}
\section{Systematic Uncertainties on Inference Results}
\label{app:systematics}
\setcounter{figure}{0}

\Needspace{4\baselineskip}
\subsection{Concentration--Mass Relation}
\label{app:concmass}

\begin{figure*}[p]
\centering
    \includegraphics[width=0.98\textwidth]{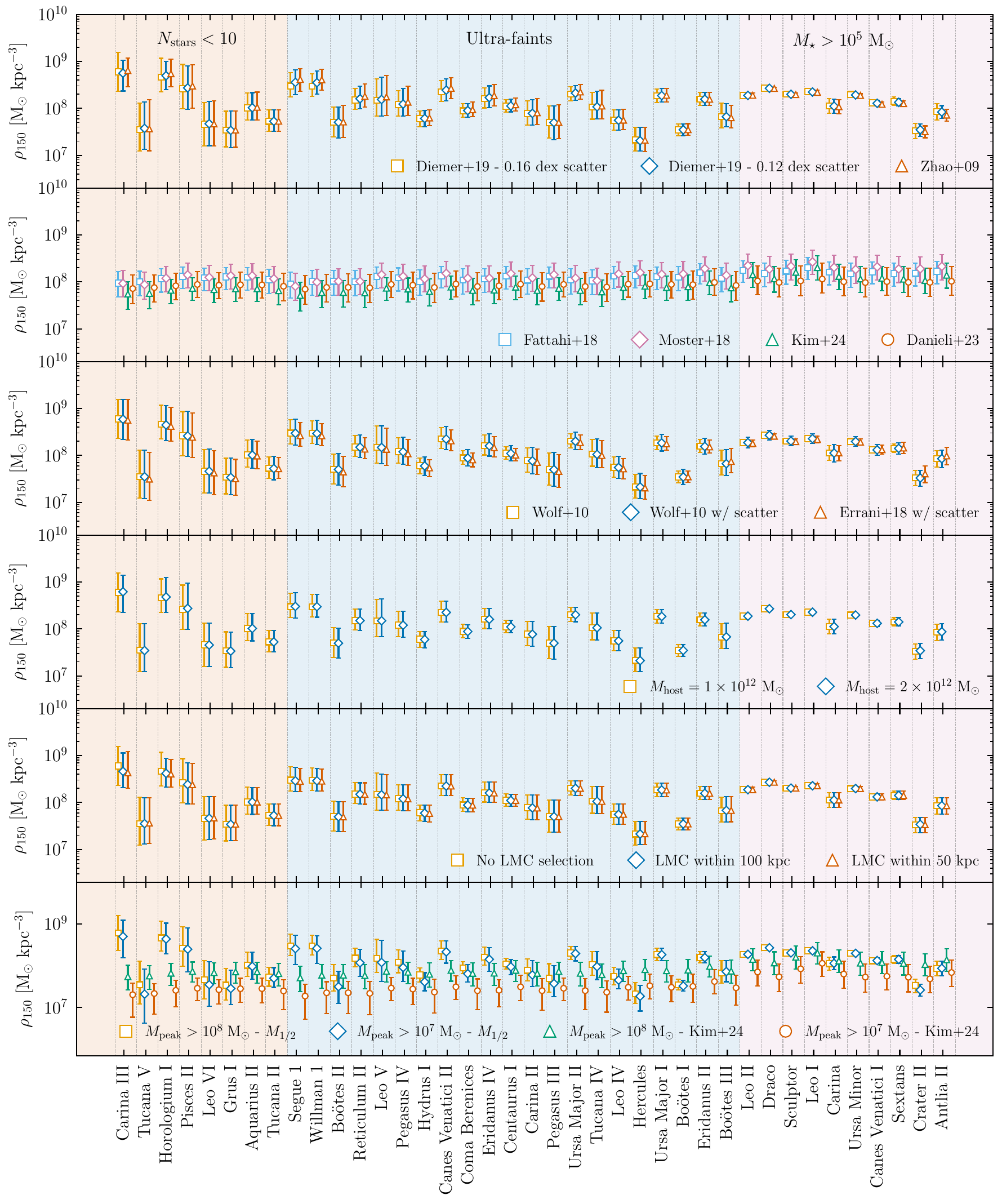}
    \caption{Comparison of inferred satellite $\rho_{150}$ values for each source of systematic uncertainty considered in this work. Dwarf galaxies follow the same ordering as \autoref{fig:multipanel}. With the exception of the second and bottom panels, which include \Mstar{} weights, all densities are inferred using \Mhalf{} weights. Unless otherwise specified, all results are computed using the Fiducial mass floor \SatGen{} run. See text for more details on each panel.}
    \label{fig:systematics}
\end{figure*}

The central densities of the \SatGen{} halos depends on the choice of concentration--mass (\concmass{}) relation, and thus different \concmass{} relations can in principle yield different inferences. To check this, the top panel of \autoref{fig:systematics} compares the \Mhalf{}-inference results for $\rho_{150}$ using three different \concmass{} relations: 
\begin{itemize}
\item The light orange squares reproduce the main-body results, which use the model of~\citet{Diemer:2018vmz} with a scatter of 0.16~dex. \citet{Diemer:2018vmz} construct a \concmass{} model analytically by describing the relation between the height of primordial density fluctuation peaks and the resulting virialized halo's concentration. The analyticity of this argument allows it to extend to very small halos, beyond the $\Mvir \gtrsim 10^{11}$~\Msun{} resolution of the $N$-body simulations used to calibrate the relation.

\item The dark blue diamonds show the results for the same \concmass{} relation, but with 0.12~dex scatter. This choice is motivated by~\citet{Bhattacharya:2011vr}, who find an intrinsic scatter in the \concmass{} relation of 0.12~dex for relaxed halos and 0.16~dex including unrelaxed halos. 

\item The dark orange triangles in the panel show the results for the \concmass{} relation from~\citet{Zhao:2008wd}.  This model computes satellite concentrations based on $t_4$, the time of formation of 4\% of the halo mass, as
\begin{equation}
    c(t) = 4 \times \left[ 1 + \left(\frac{t}{3.75t_{4}}\right)^{8.4} \right]^{1/8} \,,
\end{equation}
which was calibrated to $N$-body simulations of halos with $\Mvir \gtrsim 10^{10}$~\Msun{}. Under this model, we can directly compute the infall concentrations from the satellite assembly histories in \SatGen{}. The Zhao model generates a scatter of $\osim0.12\text{--}0.14$ dex.
\end{itemize}
There are generally minimal differences in the \Mhalf{}-inferred $\rho_{150}$ values between the three scenarios. The results for the Zhao model are typically shifted less than 20\% with respect to the Diemer 0.16-dex-scatter model, with the exception of Segue~1 and Willman~1, which are shifted upward by $\osim40$\%. Similarly, the results for the Diemer 0.12~dex--scatter model are shifted less than 10\% relative to the 0.16~dex--scatter model, with the same two exceptions seeing upward  shifts of $\osim20$\%.
We also repeat the procedure for the \Mstar{} inferences (not shown in figure) where the Zhao model gives systematically $\osim10\%$ and $\osim20\%$ lower results for the \Fattahi{} and \Kim{} cases, respectively. The two Diemer models agree to within $5\%$ across all dwarfs under both \Mstar{} inferences.

We restrict our comparison to these three models and do not consider \concmass{} relations based on power-law fits to simulations~\citep[e.g.,][]{Dutton:2014xda}. These fits can predict unphysically large concentrations at masses below the resolution limit of the simulations to which they are calibrated~\citep{Correa:2015dva}.

\Needspace{4\baselineskip}
\subsection{Stellar-to-Halo Mass Relations}
\label{app:SHMR}

In the main body, we bracket the uncertainty on the stellar-to-halo mass relation~(SHMR) by considering inference results from the \Fattahi{} and \Kim{} relations, which are based on studies of low-mass galaxies ($\Mvir \lesssim 10^{10}~\Msun$). Here, we provide some further details on these two relations and compare to results from two other SHMRs:~\citet[][hereafter \Moster{}]{2018MNRAS.477.1822M} and~\citet[][hereafter \Danieli{}]{2023ApJ...956....6D}.
The second panel of \autoref{fig:systematics} compares the \Mstar{}-inferred $\rho_{150}$ values for these four different SHMRs:

\begin{itemize}
\item The \Fattahi{} inference results are indicated by the light blue squares. This SHMR is constructed by fitting to central (non-satellite) galaxies in the A Project Of Simulating The Local Environment~(APOSTLE) hydrodynamical simulations~\citep{Sawala:2015cdf}. \Fattahi{} finds that the baryonic mass of satellite galaxies falls off rapidly below $\vmax \sim20$~\kms{}, with the low-mass end of the SHMR well-described by an exponential function of $\vmax$. As prescribed by~\Fattahi{}, we use $\vmax$ to assign the stellar masses within \SatGen{}. 

\item The \Moster{} inference results are indicated by the magenta diamonds. \Moster{} develop an empirical model that integrates star formation rates over DM halo assembly histories from $N$-body simulations and constrain the parameters of this model by fitting to galaxy surveys. However, this work considers higher-mass halos ($\Mvir{}> 10^{11}~\Msun{}$) and has not been validated for the lower-mass systems that are the focus of this work. 

\item The \Kim{} inference results are indicated by the green triangles. The authors of~\Kim{} take an alternative approach and first construct a relation between mean star formation rate, $\langle \mathrm{SFR} \rangle$, and $\vmax$. They then use this $\langle \mathrm{SFR} \rangle\text{--}\vmax$ relation to model star formation histories as a function of halo assembly history and find a power-law SHMR with a scatter that grows with decreasing halo mass. They also find that the scatter in \Mstar{} is driven by the assembly histories of low-mass dwarfs and the quenching of star formation by reionization. We adopt their analytic fit to the SHMR as a function of $M_{200}$, the mass corresponding to a virial overdensity of $\Delta=200$. 

\item The \Danieli{} inference results are indicated by the dark orange circles. This relation is inferred through a comparison of \SatGen{} runs to the Exploration of Local VolumE Satellites~(ELVES) Survey of satellites~\citep{Carlsten_2022}. They forward model a parameterized SHMR, observational selections, and observational uncertainties through \SatGen{} to simulate mock ELVES surveys, and then they find the best-fit SHMR parameters. 

\end{itemize}
Among the ultra-faint dwarfs, \Fattahi{} infers systematically larger densities than \Kim{}. This occurs because the exponential dropoff in \Fattahi{} sets a cutoff in the halo masses that can host galaxies at a few times $10^8~\Msun{}$, while the power law of \Kim{} can continue to place galaxies into the lowest-mass halos in \SatGen{}. \Danieli{} infers densities between the results of \Fattahi{} and \Kim{} for the ultra-faints, but infers the lowest densities for the $\Mstar>10^5~\Msun{}$ dwarfs. \Moster{} infers systematically larger densities than the other three relations across all but the faintest dwarf galaxies. We additionally check the inference results using other SHMRs calibrated to high-mass halos~($M_{\rm vir}>10^{10}~\Msun{}$), namely~\citet{Behroozi:2012iw,Moster:2012fv,2017MNRAS.470..651R,Behroozi:2019kql}; and \citet{Munshi:2017czb}, and find that the results are contained within the uncertainties of the four models discussed in detail here.

\Needspace{4\baselineskip}
\subsection{Dynamical Mass Estimators}
\label{app:mass_estimators}

The inference procedure outlined in the main body of this work requires the use of accurate dynamical mass estimators for dwarf galaxies. Each dynamical mass estimator considered in this appendix minimizes a function of the form $M(\lambda\times r_{1/2}) = \mu \times (\lambda \times r_{1/2}) \sigmaLOS^2 G^{-1}$ (where $r_{1/2}$ is the half-light radius, $\sigmaLOS$ is the luminosity-weighted LOS velocity dispersion, $G$ is the gravitational constant) with respect to a specific source of uncertainty---the parameters $\mu$ and $\lambda$ are chosen to minimize the relevant uncertainty. The third panel of \autoref{fig:systematics} compares \Mhalf{}-inferred $\rho_{150}$ values for different choices of dynamical mass estimator and associated scatter:
\begin{itemize}
    \item The \Wolf{} results are indicated by the light orange squares. \Wolf{} computes the mass at $r_{1/2}$ as an approximation to the radius that is minimally sensitive to velocity anisotropies. The authors of that work then use the spherical Jeans equation to derive the corresponding mass estimate. 
    \item The dark blue diamonds show the \Wolf{} results with 23\% scatter added in quadrature to the propagated measurement uncertainties. 
    \citet{Campbell:2016vkb} quantify this scatter by considering dwarf galaxies from the APOSTLE simulations~\citep{Sawala:2015cdf} and identify variations in the 3D shapes of stellar profiles as the source of the scatter.
    \item The dark orange triangles show the results for the \citet[][hereafter \Errani{}]{Errani_2018} mass estimator.  For this model,  $M(<1.8 R_h)\approx 3.5 \times 1.8 R_h \sigmaLOS^2 G^{-1}$, where $R_h$ is the 2D half-light radius. This estimator is derived from the virial theorem, minimizing the uncertainties due to ignorance of the inner slope of the DM density profile and the ratio of scales between the stellar profile and the DM density profile, $R_h/\rmax$. The intrinsic scatter on this mass estimator is $\osim10\%$. 
\end{itemize}
For the ultra-faint dwarf galaxies, the choice of mass estimator and scatter is subdominant to posterior inference uncertainties. On the other hand, for many of the high-luminosity dwarf galaxies, the scatter in the dynamical mass estimator is the dominant source of uncertainty. This effect is largest for Draco, for which the uncertainty on the inferred $\rho_{150}$ increases by a factor of $\osim4$ upon adding the 23\% scatter to the \Wolf{} estimator. 

\Needspace{4\baselineskip}
\subsection{Host Halo Mass}
\label{app:host_mass}

This subsection compares the $\rho_{150}$ inference for two sets of \SatGen{} runs with different host halo masses: $10^{12}~\Msun{}$ (i.e., the Fiducial run) and $2\times10^{12}~\Msun{}$. The satellite statistics between the two hosts are almost identical within $\osim50$~kpc, beyond which the more massive host has a greater number of satellites. At the $M_{\rm host}=10^{12}~\Msun{}$ virial radius~($\osim 260$~kpc), the more massive host contains a factor of two more satellites. 

The fourth panel of \autoref{fig:systematics} compares the \Mhalf{}-inferred central densities for the two sets of \SatGen{} runs. The $M_{\rm host}=10^{12}~\Msun{}$ results are indicated by the light orange squares, while the $M_{\rm host}=2\times10^{12}~\Msun{}$ are indicated by the dark blue diamonds. The inference results for the two host masses are consistent across all dwarf galaxies. We additionally verify that the \Mstar{}-inferred $\rho_{150}$ values for \Fattahi{} and \Kim{} are also unchanged between these two host masses.

\Needspace{4\baselineskip}
\subsection{LMC Selection}
\label{app:lmc_selection}

\begin{figure}
\centering
   \includegraphics[width=0.49\textwidth]{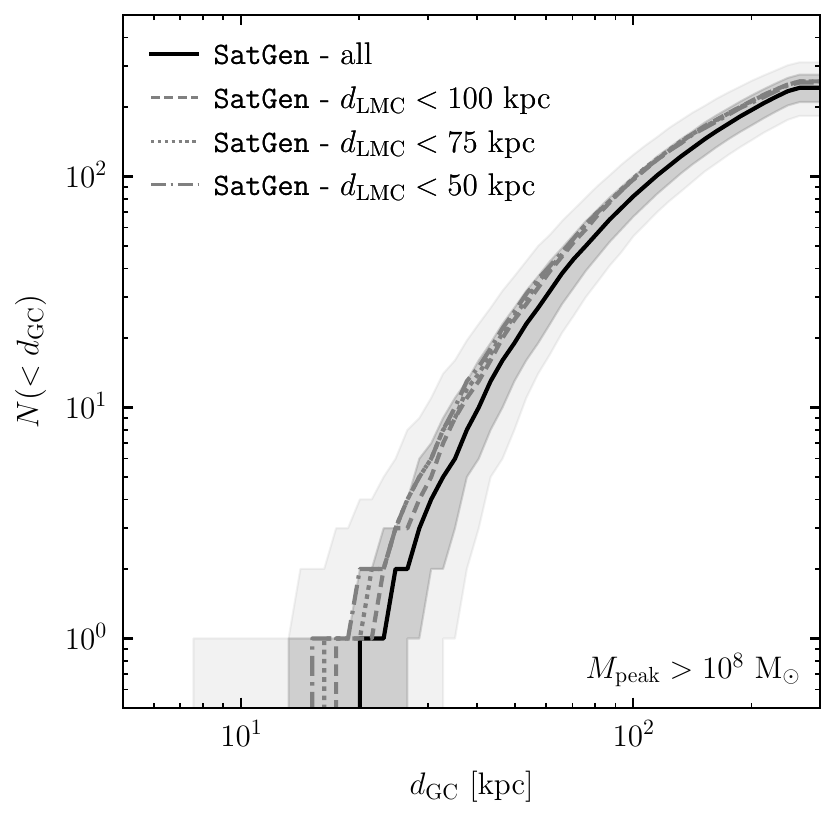}
   \caption{The number of satellites within a given galactocentric distance, $d_{\rm GC}$, of the host MW. The \SatGen{} median without any LMC selection is shown by the solid black line, along with 68\% and 95\% containment bands. The gray dashed, dotted, and dash-dotted lines show the median satellite numbers for \SatGen{} hosts with an LMC analog located within 100, 75, and 50~kpc, respectively, of the host. 
   Selecting for an LMC analog increases satellite statistics within 50~kpc by up to 40\% relative to the full \SatGen{} population, and closer LMCs lead to a greater number of satellites within 50~kpc.}
   \label{fig:distance_stats}
\end{figure}

This subsection summarizes how satellite properties and population statistics in \SatGen{} depend on the presence of an LMC analog, defined as a satellite with $\Mpeak{} > 10^{11}~\Msun{}$. The LMC analogs are, upon infall, surrounded by their own system of satellites due to the hierarchical nature of DM halo formation. These satellites may be released to the MW host, especially as the LMC analog approaches pericenter. To check for this, \autoref{fig:distance_stats} shows the number of satellite halos in the Fiducial \SatGen{} run within a given galactocentric distance, $d_{\rm GC}$, for different selections on the distance of the LMC analog, $d_{\rm LMC}$.  The medians for $d_{\rm LMC} < 50, 75,$ and $100$~kpc are shown by the dot-dashed, dotted, and dashed gray lines, respectively.  For comparison, we also show the \SatGen{} median without selecting on an LMC in solid black, along with the 68\% and 95\% containment bands. 
Requiring an LMC analog increases the satellite statistics within 50~kpc of the host by up to 40\% relative to the full \SatGen{} population, with closer LMCs leading to slightly more satellites in this range. 

As these additional satellites may have experienced pre-infall tidal stripping while members of the LMC system, we also check that the systems with LMC analogs produce similar inferred profiles for the observed MW dwarfs. The fifth panel of \autoref{fig:systematics} compares \Mhalf{}-inferred $\rho_{150}$ values  without an LMC analog~(light orange squares) to those with an LMC analog located within $50$ and $100$~kpc (dark orange triangles and dark blue diamonds, respectively). For the latter two cases, the LMC-mass subhalo is removed from the satellite population used for the inference procedure. As seen in \autoref{fig:systematics}, the LMC selection has minimal impact on the inference results for most of the dwarf galaxies, with the exception of the dwarf galaxies whose $\rho_{150}$ posteriors extend above $5\times10^8~\Msun{}~{\rm kpc}^{-3}$, namely Carina III, Horologium I, and Pisces II. The LMC selection drives a slight shift downwards in the inferred $\rho_{150}$ for these dwarf galaxies since the LMC-mass halos themselves are removed from the \SatGen{} distribution. Such massive halos are necessary to support the upper end of the velocity dispersion distributions of these dwarfs. 

We also verify that the \Mstar{}-inferred $\rho_{150}$ values for both the \Fattahi{} and \Kim{} SHMRs are unaffected by the different LMC selections.

\Needspace{4\baselineskip}
\subsection{Infall Mass Threshold}
\label{app:mass_floor}

\begin{figure*}[p]
    \includegraphics[width=0.49\textwidth]{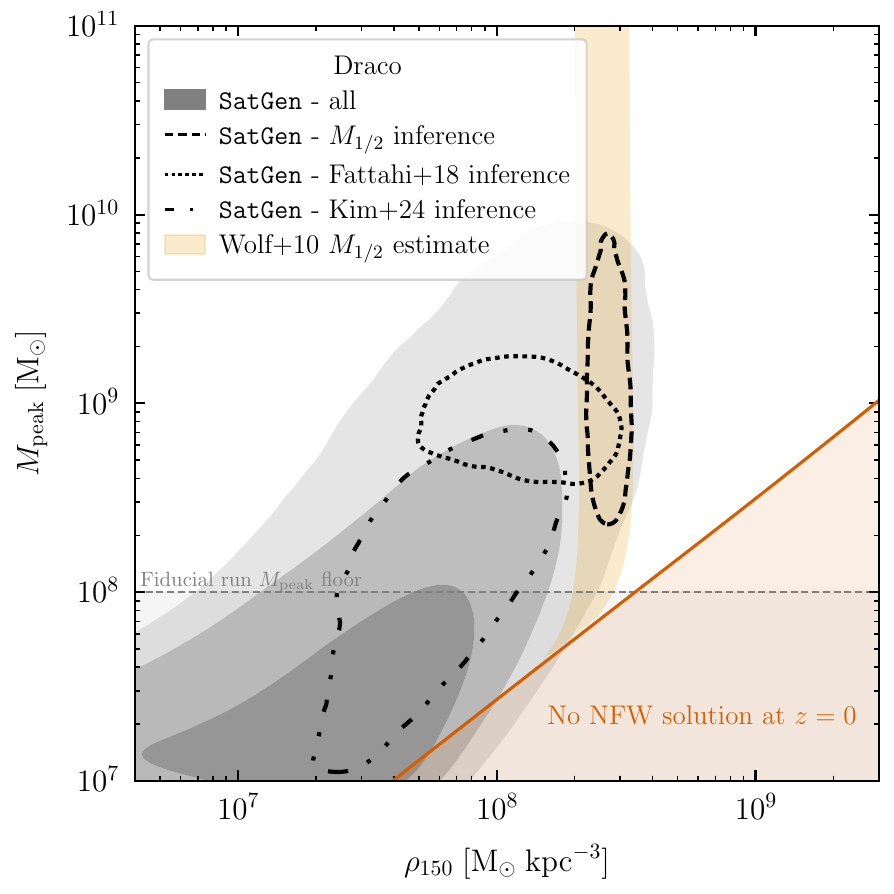}
    \includegraphics[width=0.49\textwidth]{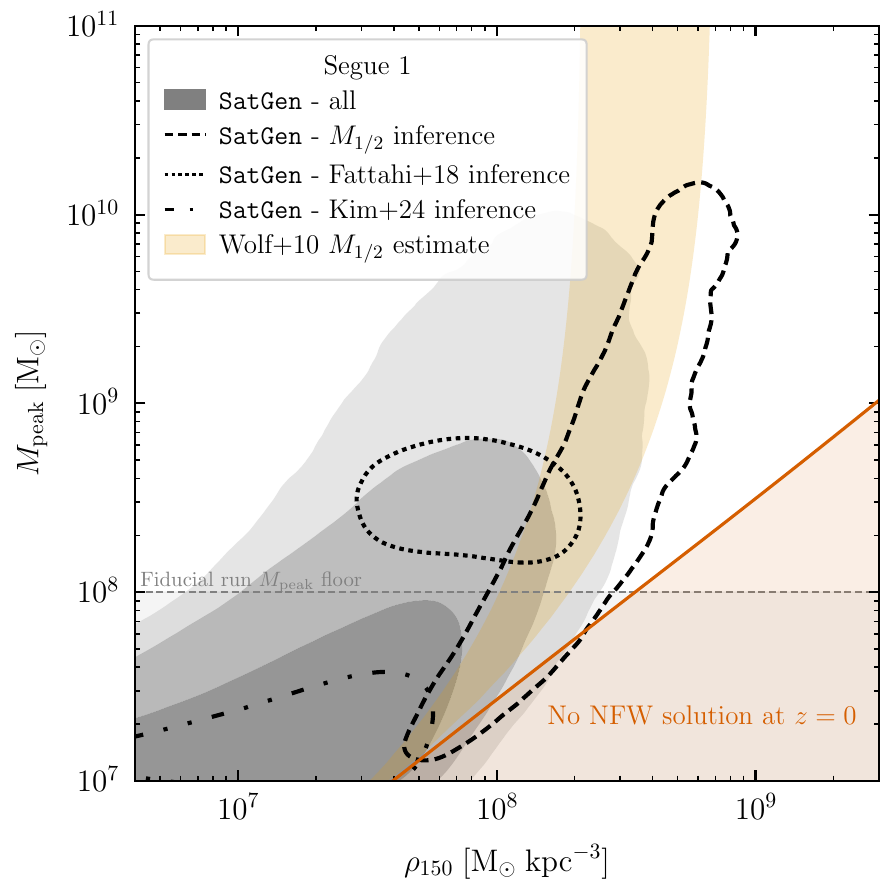}
    \par
    \centering
    \includegraphics[width=0.49\textwidth]{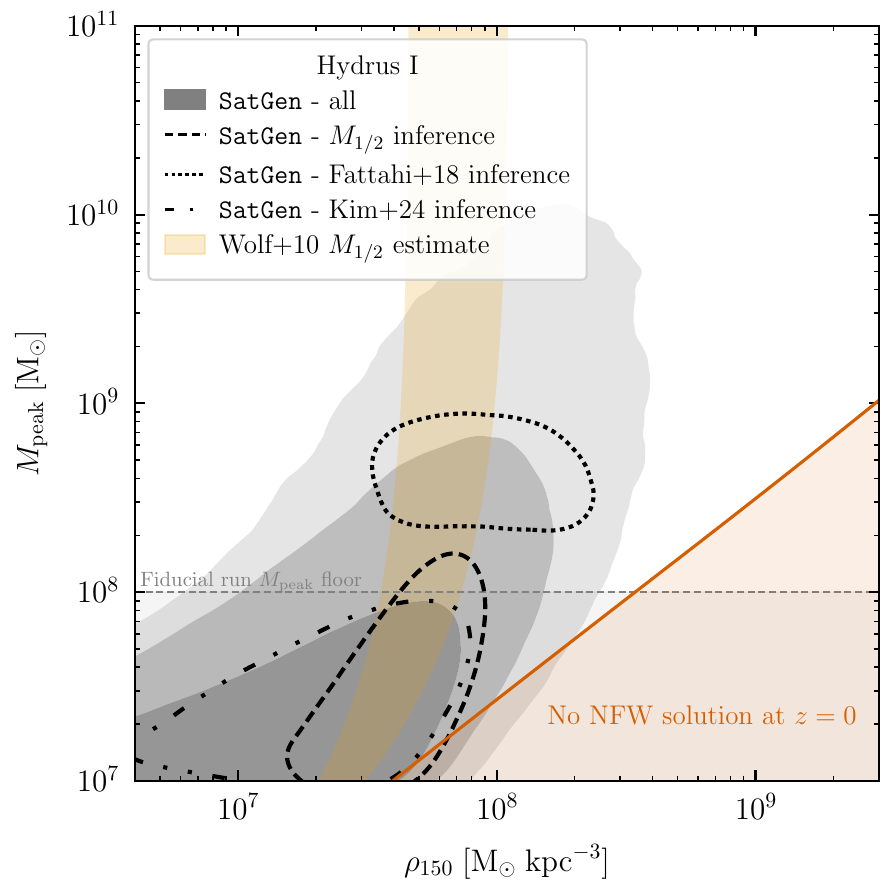}
    \caption{Inferred values for $\rho_{150}$ and \Mpeak{} for Draco~(top left), Segue~1~(top right), and Hydrus~I~(bottom). The 68, 95, and 99.5\% containment regions for the \SatGen{} satellite halos satisfying the distance selection for each dwarf are shown in gray.
    The dashed, dotted, and dash-dot-dotted black contours show the 68\% containment for the profile parameters inferred using \Mhalf{}, \Fattahi{} \Mstar{}, and \Kim{} \Mstar{} weights, respectively. The orange band shows the locus of NFW halo parameters at $z=0$ whose enclosed mass at the dwarf's half-light radius falls within the 68\% confidence interval of the observed $\Mhalf{}$ (as computed using the \Wolf{} estimator). The region without any $z=0$ NFW solution for a given $\Mpeak{}$ and $\rho_{150}$ is shaded in dark orange. Lastly, the horizontal gray dashed line indicates the fiducial mass floor of $10^8$~\Msun{}. For dwarf galaxies where the \Wolf{} \Mhalf{} band overlaps with the high-density region of the \SatGen{} distribution below the $\Mpeak{}=10^8~\Msun{}$ mass floor, the \Mhalf{}-inference contour expands or shifts to lower masses when decreasing the mass floor.
    }
    \label{fig:rho150_vs_Mpeak}
\end{figure*}

The Fiducial \SatGen{} run that is used for the primary analyses in this work has an initial mass floor of $\Mpeak=10^8~\Msun$. This step-function threshold loosely reflects the threshold for galaxy formation. In reality, though, the probability of galaxy formation is described by some smooth turn-on, generally thought to occur between $10^7$ and $10^8~\Msun{}$~\citep[see, e.g.,][]{Bullock:2000wn,DES:2019ltu,2022MNRAS.516.3944M,Ahvazi:2024txq,2024arXiv240815214K}. Because the subhalo mass function is monotonically decreasing with mass, lowering the infall mass threshold may strongly affect the cosmological prior for the inference procedure. This appendix explores the effects of lowering the mass threshold in \SatGen{}.

To bracket the uncertainty on the probability function of galaxy formation, we perform a dedicated \SatGen{} run with an infall mass threshold of $\Mpeak{}=10^7~\Msun{}$ (see \autoref{tab:simulations}) and repeat the inference procedure.  \autoref{fig:rho150_vs_Mpeak}  illustrates the 2D $\rho_{150}-\Mpeak$ inference priors and posteriors from this run for Draco, Segue~1, and Hydrus~I. The gray, filled-in contours indicate the 68, 95, and 99.5\% containment of \SatGen{} halos that pass the distance selection for each dwarf. The black dashed, dotted, and dash-dot-dotted contours show the 68\% containment contours for the \Mhalf{}-, \Fattahi{} \Mstar{}-, and \Kim{} \Mstar{}-inference results, respectively. The light orange bands show the locus of $z=0$ NFW parameters whose enclosed mass at the half-light radius falls within the 68\% confidence interval of the observed $\Mhalf{}$. At fixed mass and redshift, there is an upper bound on the NFW halo density at 150~pc; the region above this bound is shaded in dark orange. \SatGen{} halos beyond this threshold must fall in at earlier redshifts. Finally, the horizontal gray dashed line indicates the fiducial $10^8~\Msun{}$ mass floor.

As expected, the majority of halos in the $10^7~\Msun{}$ \SatGen{} run fall below the mass floor of the Fiducial run. For a dwarf like Hydrus~I, where the light orange band passes through the high-density region of the \SatGen{} distribution, the \Mhalf{}-inference contour is pushed to $\Mpeak{} < 10^8~\Msun{}$, lowering the \Mhalf{}-inferred $\rho_{150}$.  In contrast, for a dwarf like Draco, the \Mhalf{} inference contour is already resolved above the fiducial mass floor because the light orange band passes through the \SatGen{} distribution at higher densities.  For dwarfs like Segue~1, the light orange band extends below the fiducial mass floor, slightly lowering the \Mhalf{}-inferred densities once the mass floor is lowered. 

For the three dwarfs highlighted in \autoref{fig:rho150_vs_Mpeak}, the \Fattahi{} inference is fully resolved above the fiducial mass floor given its exponential cutoff at a few times $10^8~\Msun{}$. However, given its power-law behavior and increasing scatter in stellar mass at lower halo masses, the \Kim{} inference extends below the fiducial mass floor for Draco and falls fully beneath it for Hydrus~I and Segue~1.  We therefore expect to see more significant shifts in the inferred $\rho_{150}$ values for this SHMR relation in the $\Mpeak{}=10^7~\Msun{}$ \SatGen{} run.

The bottom panel of \autoref{fig:systematics} shows the \Mhalf{}-inference results for the $10^8$~($10^7$)~\Msun{} mass floor in light orange squares~(dark blue diamonds) for all the dwarf galaxies considered. The inference results for the $\Mstar>10^5~\Msun$ dwarfs are left unchanged, with the exception of Crater~II. For most UFDs, there is a slight shift downwards in the density inferred with the lower-mass-floor runs. The largest shift occurs for Tucana~V, whose median inferred density shifts downwards by a factor of $\osim 1.7$. Among the highest-density dwarfs (e.g., Carina~III, Horologium~I, Pisces II, Segue~1, and Willman~1), the largest shift in the $\rho_{150}$ median is by a factor of $\osim 1.2$ downwards for Carina~III. 

We also checked the results for the \Fattahi{} and \Kim{} \Mstar{} inference. The latter are included in the bottom panel of \autoref{fig:systematics} for the $10^8$~($10^7$)~\Msun{} mass floor in green triangles~(dark orange circles). For the \Kim{} inference, there is a systematic downward shift across all dwarfs in the median inferred density by a factor of $\osim2$--3 when lowering the mass floor. For the \Fattahi{} inference, which is not shown in the figure, there is a smaller systematic downward shift by a factor of $\osim1.1\text{--}1.5$. These downward shifts in the \Mstar{} inference broadly maintain the discrepancies with the \Mhalf{} inference at the level of individual dwarf galaxies.
For the case of \Fattahi{}, the difference between the \Mhalf{} and \Mstar{} inference on $\rho_{150}$ never shrinks by more than 0.1~dex for any dwarf in the sample.  For the case of \Kim{}, the difference shrinks by more than 0.1~dex for Antlia~II (0.11 dex), Bo\"otes~I (0.35 dex), Crater~II (0.24 dex), Grus~I (0.32 dex), Hercules (0.33 dex), and Tucana~V (0.20 dex).

Importantly, the comparison presented in \autoref{sec:vcirc_rhalf} is between the $V_{\rm circ}(r)$ power law extracted from the data and \Mstar{}-expected power-law. The former is independent of the inference procedure, while for monotonically increasing SHMRs, we expect the latter to be roughly bounded from below by the NFW power law. In fact, lowering the mass floor lowers the \Mstar{}-inferred density for the least luminous satellites. This steepens the \Mstar{}-inferred power-law and increases the discrepancy observed in \autoref{sec:vcirc_rhalf}.

\Needspace{4\baselineskip}
\section{Tidal Stripping Effects}
\label{app:tidal_stripping}
\setcounter{figure}{0}

\begin{figure}
\centering
    \includegraphics[width=0.49\textwidth]{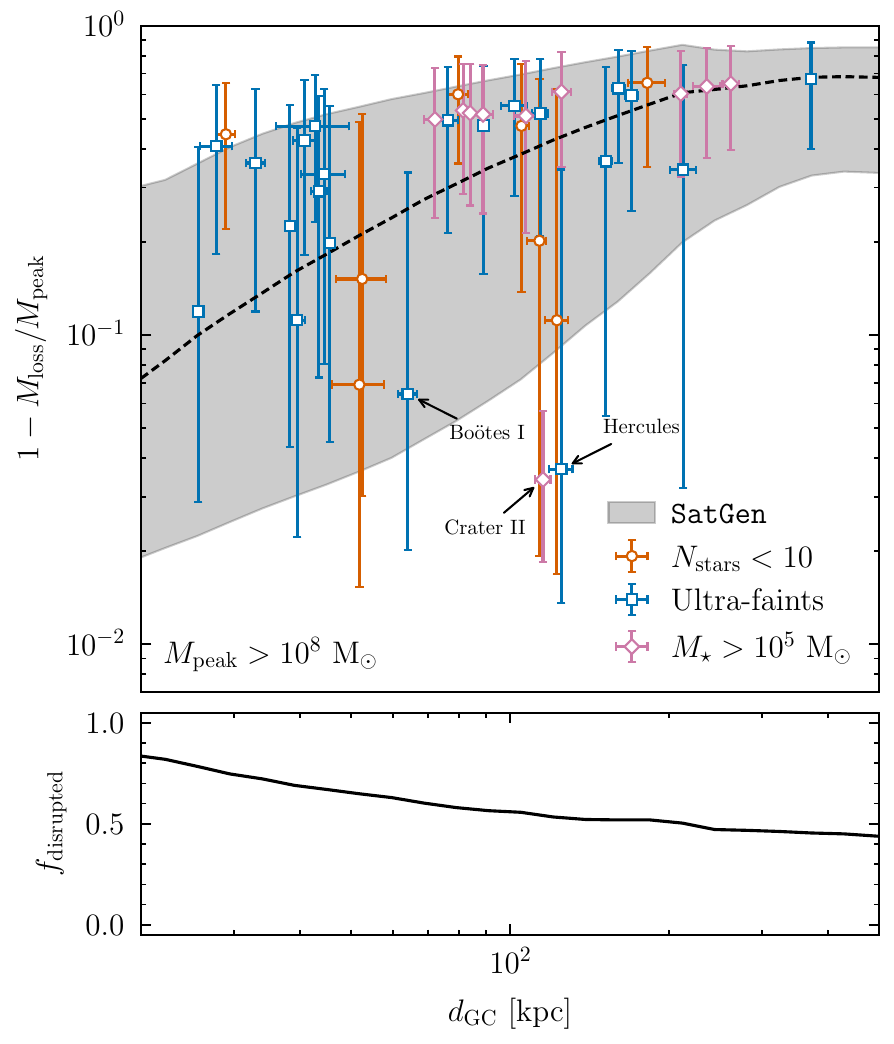}
    \caption{Top: The fraction of a satellite's mass retained after infall as a function of its $z=0$ galactocentric distance, $d_{\rm GC}$. We show the median (dashed black line) and 68\% containment band (gray) for the \SatGen{} population.  For each observed MW dwarf, the median and 68\% containment band is also provided; the dwarfs are divided into the categories discussed in \autoref{sec:data} ($N_{\rm stars} < 10$ as dark orange circles, Ultra-faints as dark blue squares, $\Mstar{}>10^5~\Msun{}$ as magenta diamonds). Bottom: The fraction of \SatGen{} subhalos that have lost at least 99\% of their infall mass by $z=0$ as a function of $d_{\rm GC}$.
    }
    \label{fig:tidal_stripping}
\end{figure} 

A key modeling feature of \SatGen{} is its ability to resolve tidal stripping effects on satellite density profiles. The upper panel of \autoref{fig:tidal_stripping} shows the fraction of infall halo mass retained by the \SatGen{} satellites as a function of $z=0$ galactocentric distance, with the caveat that we only consider satellites that retain at least 1\% of their initial halo mass. The median is indicated by the dashed-black line and the 68\% containment by the gray band.  The lower panel of \autoref{fig:tidal_stripping} shows the fraction of satellites at each $z=0$ galactocentric distance in \SatGen{} that have fallen below the 1\% threshold ($f_{\rm disrupted}$). As expected, the tidal mass loss increases at smaller distances at the population level.

The upper panel of \autoref{fig:tidal_stripping} additionally shows the retained mass fraction of MW satellites, inferred using \Mhalf{} weights. The MW satellites generally follow the trend in \SatGen{}, with mass loss increasing towards smaller galactocentric distances. The figure highlights the three dwarf galaxies---Crater~II, Hercules, and Bo\"otes I---with the lowest median retained mass fraction. Crater~II retains less than 10\% of its infall mass; \SatGen{} requires significant tidal stripping to reproduce its observed velocity dispersion. Recent dedicated studies of Crater~II have indeed found evidence for associated tidal streams~\citep{coppi2024lasillaquestrrlyrae, vivas2025crateriidisrupting}. Similarly, the median inferred mass losses for Hercules and Bo\"otes I are greater than 90\%. Studies of Hercules find an elongated shape along with RR Lyrae member stars outside its tidal radius, consistent with tidal disruption effects~\citep{Deason:2012au,Garling_2018,Ou:2024kto}. 
For Bo\"otes I, recent work finds a velocity gradient and ellipticity consistent with tidal disruption~\citep{Longeard_2022, sandford2025chemodynamicsbootesis5revised}.

\Needspace{4\baselineskip}
\section{Comparison of LVDB and KDSA}
\label{app:lvdb_keckdeimos}
\setcounter{figure}{0}

\begin{figure}
    \centering
    \includegraphics[width=0.49\textwidth]{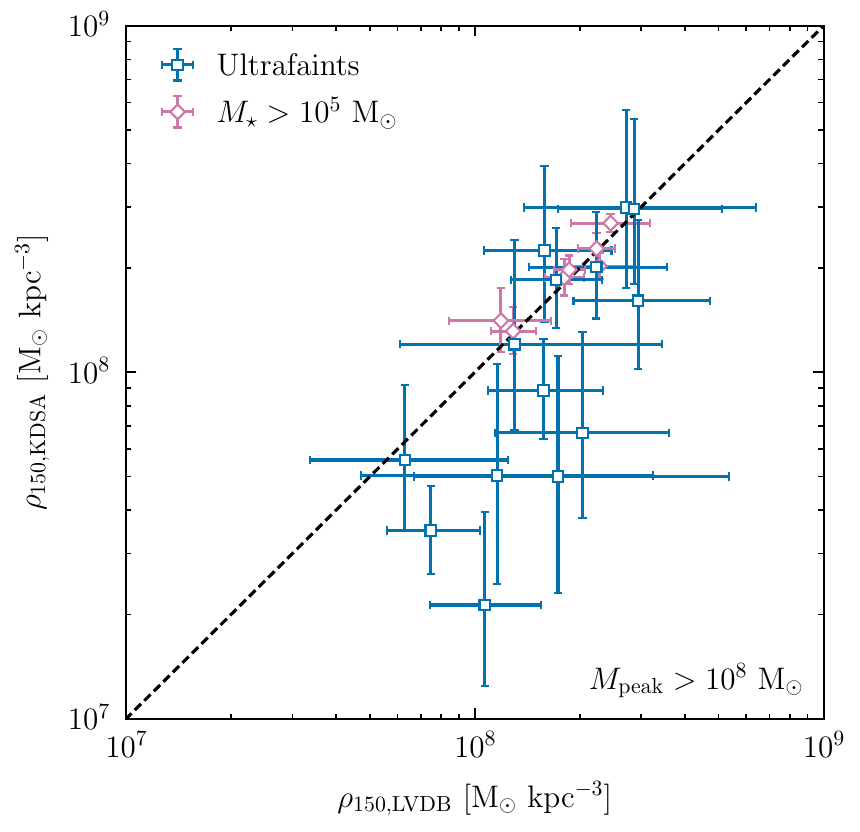}
    \caption{The \Mhalf{}-inferred $\rho_{150}$ values for the 22 dwarf galaxies overlapping between the KDSA, LVDB, and our work. The $x$-axis shows results using the LVDB velocity dispersions and the $y$-axis shows results using the updated KDSA velocity dispersions. The densities for the Ultra-faints category are shown as dark blue squares, while those for the $\Mstar{}>10^5~\Msun{}$ category are shown as magenta diamonds. 
    }
    \label{fig:lvdb_vs_kdsa}
\end{figure}

The reanalysis of Keck/DEIMOS data from~\citet{geha2026keckdeimosstellararchivei} and \citet{geha2026keckdeimosstellararchiveii} finds consistent---though generally lower---velocity dispersions for many dwarf galaxies compared to prior literature results; five galaxies have KDSA velocity dispersions at least $1\sigma$ below the LVDB dispersions. Here, we compare the impact of these shifts on our inference results. \autoref{fig:lvdb_vs_kdsa} compares \Mhalf{}-inference results for $\rho_{150}$ using the dispersions from the KDSA and the LVDB for the overlapping subset of 22 dwarf galaxies considered in this work. By construction, the KDSA only contains dispersions for dwarf galaxies with at least 10 member stars, so \autoref{fig:lvdb_vs_kdsa} only shows the Ultra-faint and $\Mstar{}>10^5~\Msun$ categories. Most dwarf galaxies are within statistical uncertainties of the one-to-one line or have lower inferred densities with the updated KDSA dispersions. 
 These findings are consistent with the results from~\citet{geha2026keckdeimosstellararchivei} and \citet{geha2026keckdeimosstellararchiveii}.

\Needspace{4\baselineskip}
\section{Comparison to $N$-body Simulations}
\label{app:simulations}
\setcounter{figure}{0}

\begin{figure*}[p]
    \centering
    \includegraphics[width=0.49\textwidth]{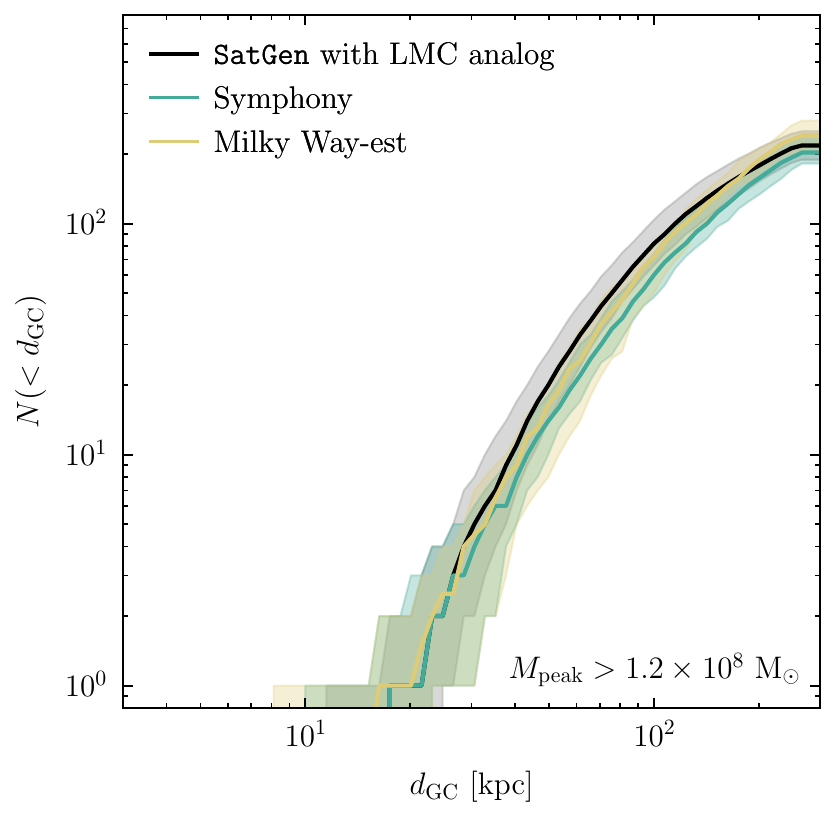}\\
    \includegraphics[width=0.49\textwidth]{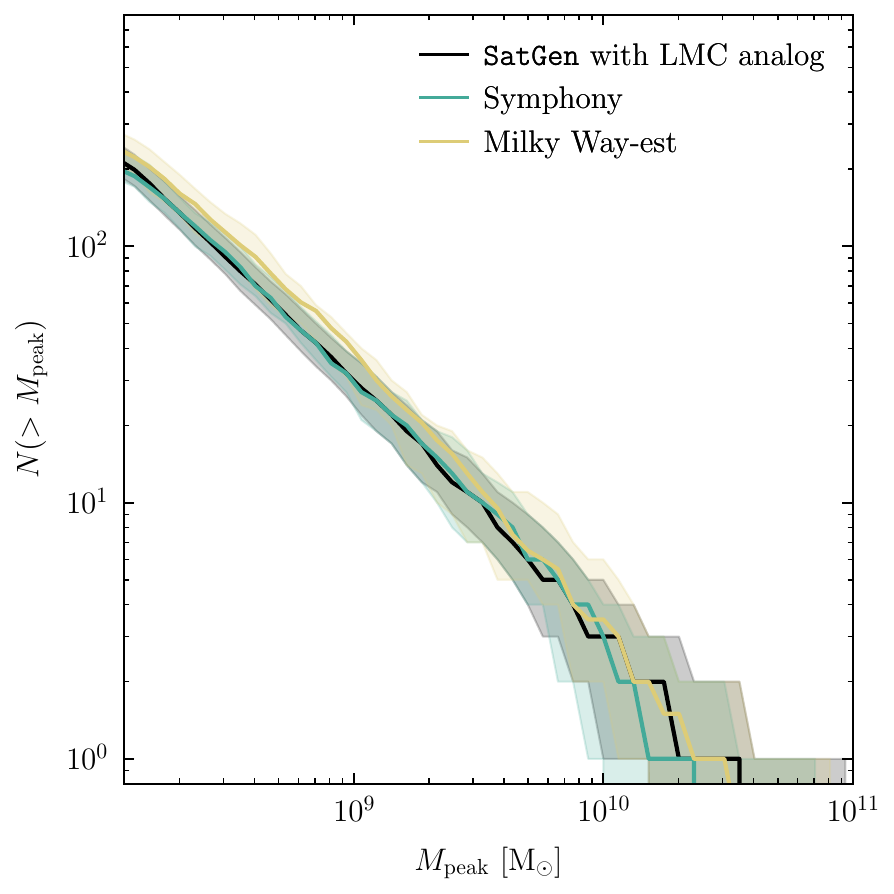}
    \includegraphics[width=0.49\textwidth]{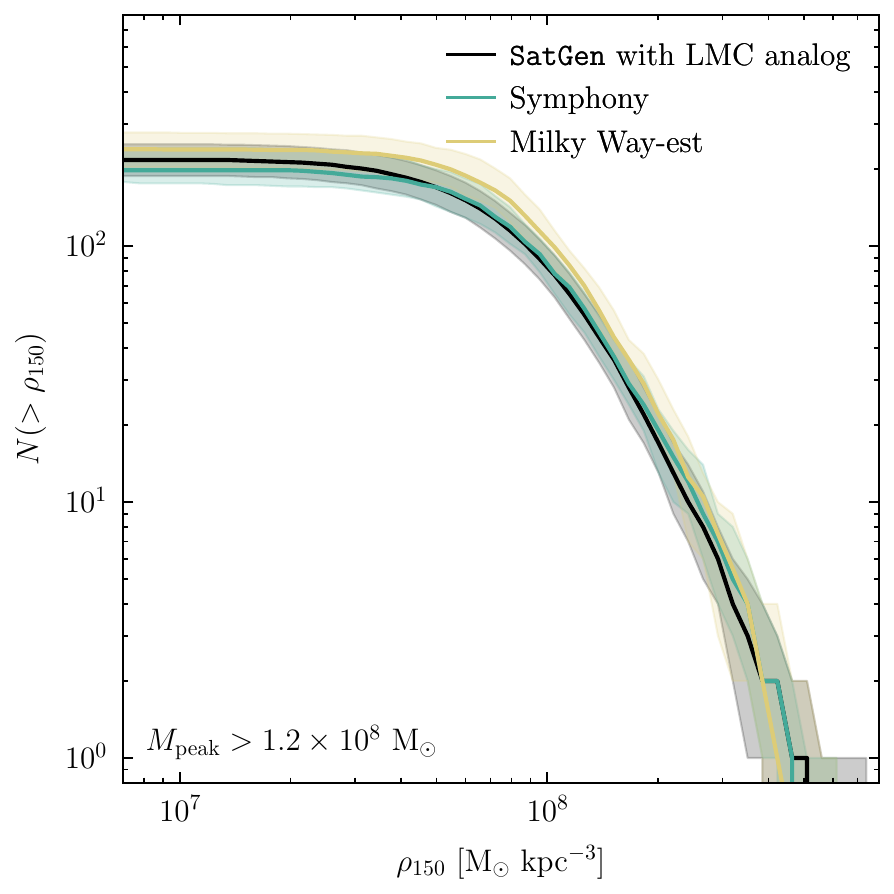}
    \caption{
    Top:~The number of satellites within a given galactocentric distance, $d_{\rm GC}$, for \SatGen{}~(gray), as well as the Milky Way-est~(yellow) and Symphony~(teal) simulations~\citep{Nadler:2022dvo,Buch:2024ssx}. In each case, the solid line indicates the median distribution and the band encompasses the 68\% containment.  Results are provided for the Fiducial \SatGen{} run, selecting for an LMC within 100~kpc of the Galactic center.  Satellites in all three samples must satisfy $\Mpeak{}>1.2\times10^8~\Msun$.
    Bottom:~Same as top panel, except for peak mass, \Mpeak{}, on the left and central density, $\rho_{150}$, on the right. Importantly, the simulations do not produce more high-mass or high--density satellites compared to \SatGen{}.
    }
    \label{fig:sims_statistics}
\end{figure*}

This appendix compares the satellite statistics of the \SatGen{} hosts with LMC analogs to dedicated CDM-only cosmological zoom-in simulations of MW-like halos: Symphony~\citep{Nadler:2022dvo} and Milky Way-est~\citep{Buch:2024ssx}. The Symphony suite consists of 45 MW-mass halos ($M_{\rm host} \sim 10^{12.09 \pm 0.02}~\Msun$). Milky Way-est is a suite of 20 hosts, simulated with the same resolution and numerical settings as Symphony, that also satisfy additional formation history constraints: (i)~the recent infall of an LMC-mass satellite that is currently located $\osim50$~kpc from the Galactic center and (ii)~an early merger with a Gaia~Sausage--Enceladus analog~\citep{Belokurov:2018vpl,Helmi:2018hnh}. The Milky Way-est host masses range from $\left(1\text{--}1.8 \right) \times 10^{12}~\Msun$. The two sets of simulations have a DM particle mass of $4\times10^5~\Msun{}$ and a force softening scale of $170$ pc h$^{-1}$. Symphony and Milky Way-est achieve a similar resolution to the Fiducial \SatGen{} run, resolving subhalos down to $1.2\times10^8~\Msun$.

The panels of \autoref{fig:sims_statistics} compare the distribution of satellite distance~(top), \Mpeak{}~(bottom left), and $\rho_{150}$~(bottom right) across the Milky Way-est (yellow), Symphony (teal), and \SatGen{} (gray) samples. The latter consists of all satellites in the Fiducial \SatGen{} run with $\Mpeak{}>1.2\times10^8~\Msun$. Additionally, we only consider \SatGen{} halos associated with hosts that have an LMC-analog satellite located within 100~kpc of the Galactic center; Milky Way-est selects for LMC analogs between 30 and 70 kpc which, following the discussion of \hyperref[app:lmc_selection]{Appendix~\ref*{app:lmc_selection}}, may increase the satellite statistics relative to the 100~kpc selection. In \SatGen{}, requiring an LMC-analog satellite within 50~kpc of the Galactic center increases the number of satellites within this region by $\osim$10\% relative to the case where there is an LMC-analog satellite within 100~kpc. Within 100~kpc of the Galactic center, the satellite statistics in both cases are within 1\% of each other.

As shown in the top panel of \autoref{fig:sims_statistics}, the Symphony and Milky Way-est subhalo number functions are consistent with \SatGen{} across all distances. The distributions of \Mpeak{} and $\rho_{150}$, also selecting for halos with $\Mpeak{}>1.2\times10^8~\Msun$, are shown in the bottom panels of \autoref{fig:sims_statistics}. There is generally excellent agreement between the \SatGen{}, Milky Way-est, and Symphony results. 
Importantly, the numerical simulations do not significantly overpredict subhalo statistics at high masses and densities relative to \SatGen{}.  

\Needspace{4\baselineskip}
\section{Analytic derivation of power-law behavior}
\label{app:powerlaw}
\setcounter{figure}{0}

The main body of this work motivates the power-law behavior of circular velocity profiles by pointing to the leading behavior of an NFW halo~\citep{Navarro:1995iw,Navarro:1996gj} at $r\lesssim r_s$, where $r_s$ is the scale radius. This would be sufficient to explain the results from the \SatGen{} realizations if the halos hosting the UFDs were all the same. However, that is not true. This appendix shows analytically why the power-law expectation holds more generally for the median \Vcirc(\rhalf). 

Given the scale radius $r_s$ and the characteristic density $\rho_s$ of an NFW density profile, the circular velocity at $x\equiv r/r_s$ is $\Vcirc^2(r)=2\pi G\rho_s r_s r g(x)$ with dimensionless acceleration $g(x)=2(\ln(1+x)-x/(1+x))/x^2$ such that $g(0)=1$. We write the expansion rate at redshift $z$ as $H(z)=H_0 E(z)$, which defines $E(z)$. Defining the halo mass  $M_\Delta$ as the mass within the radius $r_\Delta$ such that the mean density is $\Delta \times \rho_\mathrm{crit} E(z)^2$, and the concentration $c=r_\Delta/r_s$ \citep[e.g.,][]{Bryan:1997dn}, we can write the scale density and radius in terms of $M_\Delta$ and $c$ as $4\pi G \rho_s r_s=\Delta H_0^2 E(z)^2 r_\Delta/g(c)$.   
Then, we can rewrite the circular velocity as 
\begin{equation}
    \Vcirc(r) = 
    \sqrt{\frac{\Delta}{200}}
    \,E(z) 
    \sqrt{\frac{g(x)}{g(c)}}
    \, \frac{\sqrt{r_\Delta r}}{h^{-1}\mathrm{kpc}}
    \,{\rm km/s}\,.
\end{equation}
For reference, the scale radius $r_s \approx 1\ \mathrm{kpc}\,
      (M_\Delta/10^{9}M_\odot)^{1/3}\,
      (c/21)^{-1}\, (E(z)^2\Delta/200)^{-1/3} (h/0.68)^{-2/3}$, which shows that the subhalos hosting the UFDs will have scale radius around a kpc pre-infall.  

Note that we are using relations that are appropriate before tidal stripping is operative. Because we are interested in the densities in the inner regions of the halos (corresponding to UFD half-light radii smaller than about 200 pc), we will assume that the inner densities are not changed significantly by tidal stripping. This is a good approximation~\citep{Du:2024sbt}, and it allows us to use the density profiles of subhalos at infall to estimate the circular velocity at \rhalf{} of UFDs at $z=0$. 

It will be useful to pull out the redshift dependence in the median concentration, $c_\mathrm{med}(M_\Delta,z)$. Keeping this in mind, we define $F(M_\Delta,z) = E(z)^{2/3} g\left(c_\mathrm{med}(M_\Delta,z)\right)^{-1/2}$. 
Focusing on the region $x < 1$, we can Taylor expand $g(x)$ and write
\begin{equation}
  \Vcirc(r)\approx F(M_\Delta,z)\,
  \sqrt{\frac{g(c_{\rm med})}{g(c)}}\;
  \left(\frac{M_\Delta}{10^{9}M_\odot}\right)^{1/6}
  \left(\frac{\Delta}{200}\right)^{1/3}
  \left(\frac{h}{0.7}\right)^{2/3}
  \left(\frac{r}{100\ \mathrm{pc}}\right)^{1/2}
 \left[1-\frac{2}{3}x+\frac{19}{36}x^{2}+\dots\right] \mathrm{km\,s^{-1}}\,.
  \label{eq:app-vc-numeric}
\end{equation}
Note that for the halo masses of interest here ($\osim 10^{8}\text{--}10^9~\Msun{}$), $F(z)$ has only a mild dependence on redshift because the increase in $E(z)$ is canceled by the decrease in $c_\mathrm{med}$ for $z\lesssim 4$. For this halo mass range, $F$ is about 8--9 at $z=0$ for the \Diemer{} model.

In this halo mass range, we recast the concentration in terms of parameter $s$ such that $c \approx c_9(z) \left(M_\Delta/10^9~\Msun\right)^{-\gamma} 10^s$
where $c_9(z)$ is the median concentration at $M_{\rm vir}=10^9~\Msun{}$, and $s$ is drawn from a normal distribution with width about $0.16~{\rm dex}$ (log-normal scatter in the \concmass{} relation). A value of $\gamma=0.07$ is a good fit to the model in \Diemer~at low redshifts but $\gamma$ is smaller (as expected) at higher redshift where the mass dependence flattens out. Further, we note that we can approximate $\sqrt{g(c)} \approx 0.95 c^{-0.75}$, and this shows that  the halo mass dependence in $F$ is mild, approximately scaling as $M_\Delta^{-0.05}$, resulting in \Vcirc{} scaling as $M_\Delta^{0.11}$. We can also see that $\Vcirc \propto 10^{0.75 s}$ approximately, at fixed $M_\Delta$. 

A couple of illustrative cases are useful to consider. First, if \rhalf{} is strongly correlated with the virial mass, then the correlation should be such that larger \rhalf{} would be found in more massive halos~\citep{Kravtsov:2012jn}. In such a case, $V_{\rm circ}(\rhalf)$ would scale more steeply than $(\rhalf)^{0.5}$ (not including corrections provided by non-zero $x=r/r_s$ values) because of its dependence on $M_\Delta$. Second, if there is no correlation with $M_\Delta$ and $s$ (or only weak correlations), then we see that the predicted slope, $d\ln(\Vcirc)/d\ln(\rhalf)$, will be 0.5 for small $r/r_s$. In this limit, we expect the log-normal scatter in $\Vcirc{\rhalf}$ to be $0.75 \times 0.16 + 0.11 \sigma_M$~dex, where $\sigma_M$ is the scatter in $\log_{10}(M_\Delta)$ for halos that host the UFDs. For example, for a $2\sigma$ containment band of halo masses spanning $10^8\text{--}10^9~\Msun{}$, we have $\sigma_M=0.25$, which adds about 0.03~dex to the scatter and leads to an intrinsic scatter of about 0.15~dex for the $V_{\rm circ}$ predictions. Including tidal stripping of subhalos, the scatter should increase in the direction of lower circular velocity values and also decrease median $\Vcirc(\rhalf)$, but this is not a large effect because even with 90\% mass loss, the decrement in the density in the central halo is about 20\%~\citep{Du:2024sbt}. The tidal stripping effect is captured by the full analysis using \SatGen{}. 

\end{document}